\documentstyle[prb,aps,multicol,epsf]{revtex}
\input psfig

\font\ninerm=cmr9
\def\be{\begin{equation}}
\def\ee{\end{equation}}

\def\sqr#1#2{{\vcenter{\vbox{\hrule height .#2pt
      \hbox{\vrule width .#2pt height#1pt \kern#1pt
      \vrule width.#2pt}
      \hrule height.#2pt}}}}
\def\square{\mathchoice\sqr34\sqr34\sqr{2.1}3\sqr{1.5}3}

\begin{document}
\bibliographystyle{simpl1}

\title{Width and Magnetic Field Dependence of Transition Temperature
in Ultranarrow  Superconducting Wires}

\author{Robert A Smith$^1$, Beccy S Handy$^1$ and Vinay Ambegaokar$^2$}

\address{$^1$ School of Physics and Astronomy, University of
Birmingham, Edgbaston, Birmingham B15 2TT, ENGLAND}

\address{$^2$ Laboratory of Atomic and Solid State Physics, Cornell University,
Ithaca NY14583, USA}

\maketitle
\bigskip

\begin{abstract}
We calculate the transition temperature in ultranarrow superconducting
wires as a function of resistance, wire width and applied magnetic
field. We compare the results of first-order perturbation theory and
the non-perturbative resummation technique developed by Oreg and
Finkel'stein. The latter technique is found to be superior as it is
valid even in the strong disorder limit.  In both cases, the predicted
additional suppression of the transition temperature due to the
reduced dimensionality is strongly dependent upon the boundary
conditions used.  When we use the correct (zero-gradient) boundary
conditions, we find that theory and experiment are consistent,
although more experimental data is required to verify this
systematically. We calculate the magnetic field dependence of the
transition temperature for different wire widths and resistances in
the hope that this will be measured in future experiments. The
predicted results have a rich structure -- in particular we find a
dimensional crossover which can be tuned by varying either the width
of the wire or its resistance per square.
\end{abstract}

\section{Introduction}

Recent experiments on disordered superconducting
wires\cite{GMDB87,SHD93,XHD97,BLT00} have led to renewed interest in
the interplay of interaction and disorder in low-dimensional
systems. These experiments have been made possible by the development
of fabrication techniques using either stencils or carbon nanotubes,
and involve wires of width as low as $50\AA$ and thickness as low as
$10\AA$. An obvious question to ask in such systems is how the
transition temperature is affected by wire width and
thickness. Decreasing the thickness of the wire increases the normal
state resistance per square which is a measure of the amount of
disorder in the system.  This increase in disorder reduces the
transition temperature by causing the electrons to move diffusively
rather than ballistically, as discussed below.  Decreasing the width
of the wire leads to a crossover from two- to one-dimensional
behavior, where increased fluctuation effects further reduce the
transition temperature.  Ultimately the combined disorder and
fluctuation effects completely suppress superconductivity, and a
superconductor-insulator transition results\cite{HLG89}.  The effect
of dimensionality on this quantum phase transition is obviously very
interesting from a theoretical point of view.  As a first step to
understanding this transition, we will compare predictions of
transition temperature to experimental results. Unfortunately there is
very little data available for comparison because each data point
requires the fabrication of a new wire which is very costly in terms
of experimental effort. We therefore suggest the application of a
perpendicular magnetic field as a way to generate more data points for
each wire and make predictions for this effect.

The most basic model used\cite{MF82,Fink87,SRW95} consists of electrons
moving in a disordered background interacting via the Coulomb repulsion and a featureless
phonon-mediated attraction. Simple as this model is, significant difficulties
arise even when calculating the suppression of transition temperature in two-dimensional
films. These difficulties will be compounded when looking at quasi one-dimensional wires,
which are in a crossover regime, so it is most important that they are fully understood
in the simpler two-dimensional case. The basic physics involved is not in dispute, and can
be understood from the BCS-MacMillan formula for the transition temperature, $T_c$,
%%%%%%%%%%%%%%%%%%%%%%%%%%%%%%%%%%%%%%%%%%%%%%%%%%%%%%%%%%%%%%%%%%%%%%%%%%%%%%%%%%%%%%%%%%
\begin{equation}
\label{BCS}
T_c=1.13\omega_D\exp{\left(-{1\over N(0)[\gamma-\mu^*]}\right)},
\end{equation}
%%%%%%%%%%%%%%%%%%%%%%%%%%%%%%%%%%%%%%%%%%%%%%%%%%%%%%%%%%%%%%%%%%%%%%%%%%%%%%%%%%%%%%%%%%
where $\omega_D$ is the Debye frequency (a typical phonon frequency),
$N(0)$ is the electronic density of states per spin at the Fermi
surface, $\gamma$ is the featureless attraction, and $\mu^*$ is the
Coulomb pseudopotential, a measure of the Coulomb repulsion. The
disorder causes the electrons to move diffusively rather than
ballistically making them less efficient at screening the Coulomb
repulsion. This increases $\mu^*$ and decreases $N(0)$, both effects
leading to a decrease in $T_c$. We shall assume that the disorder does
not affect phonon properties such as $\omega_D$ and $\gamma$, an
assumption that is backed up by detailed calculation on at least one
model of electron-phonon interaction\cite{Sch73}.  Other
authors\cite{Bel87} have questioned this assumption, and we intend to
address this point in future work.

The first difficulty which occurs in two dimensions relates to the
low-momentum singularity of the disorder-screened Coulomb potential,
which is given by
%%%%%%%%%%%%%%%%%%%%%%%%%%%%%%%%%%%%%%%%%%%%%%%%%%%%%%%%%%%%%%%%%%%%%%%%%%%%%%%%%%%%%%%%%%
\begin{equation}
\label{VCsing}
V_C(q,\omega_m)={4\pi e^2\over q^2}{Dq^2+|\omega_m|\over Dq_{TF}^2+Dq^2+|\omega_m|}
\approx {1\over 2N(0)}{Dq^2+|\omega_m|\over Dq^2},
\end{equation}
%%%%%%%%%%%%%%%%%%%%%%%%%%%%%%%%%%%%%%%%%%%%%%%%%%%%%%%%%%%%%%%%%%%%%%%%%%%%%%%%%%%%%%%%%%
where $q_{TF}$ is the Thomas-Fermi screening wavenumber defined by
$q_{TF}^2=8\pi e^2N(0)$, $D=v_F^2\tau/2$ is the diffusion constant,
$v_F$ is the Fermi velocity, and $\tau$ is the elastic scattering
time.  This singularity leads to a strong suppression of the
one-particle density of states\cite{AAL80} which is seen
experimentally in tunneling measurements\cite{VDG89}.  Naively
inserting this density of states into Eqn. [\ref{BCS}] predicts a
large suppression of $T_c$ which is not seen experimentally. This
prediction is, in fact, incorrect due to a
cancellation\cite{Fink87,SRW95} between the various diagrammatic
contributions in perturbation theory. The apparently mysterious
cancellation turns out to be due to gauge invariance\cite{SA00} and is
thus expected to be a very robust phenomenon. The end result is that
the Coulomb interaction is effectively featureless with strength
$N(0)V_C=1/2$. The fact that we can legitimately use a featureless
interaction makes calculations much simpler.  Our first task is
therefore to check that this replacement is still valid in quasi-1D
systems.

The other difficulty is that perturbation theory cannot take us to the
superconductor-insulator transition as it is only valid for weak
disorder, and {\it ad hoc} attempts to extend its range of validity
lead to unphysical reentrance problems.  Oreg and
Finkel'stein\cite{OF99} overcame this problem by developing a
resummation technique, which has recently been shown to be related to
strong-coupling theory\cite{SS00}.  We reexamine their predictions for
wires, and find, rather surprisingly, that the problem is extremely
sensitive to the boundary conditions used. This unexpected sensitivity
leads us to check all details and approximations used in the
resummation technique. Upon satisfactorily completing the necessary
checks, we compare theory and experiment. Although good agreement is
obtained when we focus on the suppression of $T_c$ due to increasing
disorder, it is less good when we focus on the additional suppression
due to decreasing wire width. This is largely due to the limited
amount of experimental data (only about four data points for each of
four wire widths). We propose the application of a perpendicular
magnetic field as a means of generating more data with a minimum of
effort. We extend the resummation technique to include a magnetic
field, and make predictions for transition temperature as a function
of magnetic field for different wire widths and thicknesses. The most
interesting feature which emerges is a crossover from two- to
one-dimensional behavior which can be tuned by either decreasing wire
width or increasing the resistance per square. We hope that the
appropriate experiments will be carried out in the near future.

The remainder of the paper is organised in the following manner.
In section (II) we consider first-order perturbation theory in the absence of a magnetic
field. This is the simplest approach possible, and we will use it as a testing ground
for the validity of the approximations used.
Specifically we will show that it is legitimate to use a featureless Coulomb
interaction and ignore self-energy effects, but that the details of boundary conditions
are important. In section (III) we make predictions for the transition temperature using
the resummation technique, all the necessary approximations having been justified in
section (II), and then compare theory and experiment.
In section (IV) we consider the effect of an applied magnetic field on
the transition temperature. The dimensions of the wires are such that we must consider both
orbital and Zeeman (spin) effects\cite{Clo62}. As yet there is no experimental data available 
for comparison to the theory.

\section{First Order Perturbation Theory}

We will now calculate the transition temperature, $T_c$, which we
identify as the temperature at which the pair propagator,
$L(q,\Omega=0)$, diverges. The latter is defined in Fig. 1(a) in terms
of the BCS interaction, $\gamma$, and the pair polarization function,
$P(q,0)$, by
%%%%%%%%%%%%%%%%%%%%%%%%%%%%%%%%%%%%%%%%%%%%%%%%%%%%%%%%%%%%%%%%%%%%%%%%%%%%%%%%%%%%%%%% 
\begin{equation}
\label{Ldef}
L(q,0)^{-1}=\gamma^{-1}-P(q,0).
\end{equation}
%%%%%%%%%%%%%%%%%%%%%%%%%%%%%%%%%%%%%%%%%%%%%%%%%%%%%%%%%%%%%%%%%%%%%%%%%%%%%%%%%%%%%%%%
Let us first consider mean field theory, which corresponds to zeroth
order in disorder perturbation theory, for which the expansion
parameter is $1/k_F\ell$, where $k_F$ is the Fermi wavenumber and
$\ell=v_F\tau$ is the elastic mean free path. The mean field pair
polarization function is shown in Fig. 1(b), and leads to the pair
propagator
%%%%%%%%%%%%%%%%%%%%%%%%%%%%%%%%%%%%%%%%
\begin{equation}
\label{L0def}
L_0(q,0)^{-1}=N(0)\left[\ln{\left({T\over T_{c0}}\right)}
+\psi\left({1\over 2}+{Dq^2\over 4\pi T}\right)
-\psi\left({1\over 2}\right)\right].
\end{equation}
%%%%%%%%%%%%%%%%%%%%%%%%%%%%%%%%%%%%%%%%
The first-order perturbative corrections to the pair polarization function, $\delta P(q,0)$,
are shown in Fig. 1(c), and lead to a suppression in transition temperature 
%%%%%%%%%%%%%%%%%%%%%%%%%%%%%%%%%%%%%%%%%%%%%%%%%%%%%%%%%%%%%%%%%%%%%%%%%%%%%%%%%%%%%%%%
\begin{equation}
\label{Tcdef}
\ln{\left({T_c\over T_{c0}}\right)}={\delta P(q,0)\over N(0)}.
\end{equation}
%%%%%%%%%%%%%%%%%%%%%%%%%%%%%%%%%%%%%%%%%%%%%%%%%%%%%%%%%%%%%%%%%%%%%%%%%%%%%%%%%%%%%%%%
Note that we evaluate the pair propagator at non-zero external
momentum, $q$, since we will eventually consider the effect of a
magnetic field.

The five diagrams of Fig. 1(c) can be understood in physical terms:
diagrams 1 and 2 are self-energy diagrams and give corrections to the
electronic density of states, $N(0)$; diagrams 3 and 4 are
pseudopotential diagrams and give corrections to the effective Coulomb
repulsion, $\mu^*$; diagram 5 involves the effect of Coulomb
interaction on superconducting fluctuations. We will first perform the
calculations exactly, and demonstrate the cancellation of the
low-momentum singularity in the Coulomb interaction which occurs
between diagrams 1--4 and 5. We show that this is equivalent to
considering only diagrams 1--4 with a featureless Coulomb interaction
of strength $N(0)V_C=1/2$. In other words, the only purpose of diagram
5 is to remove the low-momentum singularity, as required by gauge
invariance. Finally we show that the self-energy diagrams 1 and 2 can
be ignored as they give a much smaller contribution than diagrams 3
and 4.  The contributions from the five diagrams are as follows:
%%%%%%%%%%%%%%%%%%%%%%%%%%%%%%%%%%%%%%%%
\begin{eqnarray}
\label{bubble}
\delta P_1&=&\displaystyle-4\pi N(0)T\sum_{\epsilon_l}T\sum_{\omega_m}{1\over aL}\sum_{q'}
{Dq^2+Dq^{\prime 2}+3|\epsilon_l|+|\epsilon_l+\omega_m|\over
(Dq^2+2|\epsilon_l|)^2 (Dq^{\prime
2}+|\epsilon_l|+|\epsilon_l+\omega_m|)^2}V_C(q^{\prime},|\omega_m|)
\theta(-\epsilon_l(\epsilon_l+\omega_m))
\nonumber\\
\delta P_2&=&\displaystyle 4\pi
N(0)T\sum_{\epsilon_l}T\sum_{\omega_m}{1\over aL} \sum_{q^{\prime}}
{1\over
(Dq^2+2|\epsilon_l|)^2(D(q^{\prime}+q)^2+|\epsilon_l|+|\epsilon_l+\omega_m|)}
V_C(q^{\prime},|\omega_m|)\theta(\epsilon_l(\epsilon_l+\omega_m))\nonumber\\
\delta P_3&=&\displaystyle -4\pi
N(0)T\sum_{\epsilon_l}T\sum_{\omega_m}{1\over aL} \sum_{q^{\prime}}
{Dq^2+Dq^{\prime 2}+|\epsilon_l|+|\epsilon_l+\omega_m|\over (Dq^2+2|\epsilon_l|)
(Dq^2+2|\epsilon_l+\omega_m|)(Dq^{\prime 2}+|\epsilon_l|+|\epsilon_l+\omega_m|)^2}
V_C(q^{\prime},|\omega_m|)\theta(-\epsilon_l(\epsilon_l+\omega_m))
\nonumber\\
\delta P_4&=&\displaystyle -4\pi N(0)T\sum_{\epsilon_l}T
\sum_{\omega_m} {1\over aL}\sum_{q^{\prime}}
{1\over (Dq^2+2|\epsilon_l|)(Dq^2+2|\epsilon_l+\omega_m|)
(D(q^{\prime}+q)^2+|\epsilon_l|+|\epsilon_l+\omega_m|)}
V_C(q^{\prime},|\omega_m|)\theta(\epsilon_l(\epsilon_l+\omega_m))\nonumber\\
\delta P_5&=&\displaystyle 4\pi^2N(0)^2T\sum_{\omega_m}{1\over aL}\sum_{q^{\prime}}\left[
T\sum_{\epsilon_l}\left\{
\frac{|\omega_m|}{(Dq^2+2|\epsilon_l|)(Dq^2+2|\epsilon_l+\omega_m|)
(D(q+q^{\prime})^2+|\epsilon_l|+|\epsilon_l+\omega_m|)}
\theta(\epsilon_l(\epsilon_l+\omega_m))\right.\right.\nonumber\\
&&+\left.\left.
\frac{Dq^2+|\omega_m|}{(Dq^2+2|\epsilon_l|)(Dq^2+2|\epsilon_l+\omega_m|)
(D(q+q^{\prime})^2+|\epsilon_l|+|\epsilon_l+\omega_m|)}
\theta(-\epsilon_l(\epsilon_l+\omega_m))\right\}\right]^2
V_C(q^{\prime},|\omega_m|)L(q+q^{\prime},|\omega_m|)
\end{eqnarray} 
%%%%%%%%%%%%%%%%%%%%%%%%%%%%%%%%%%%%%%%%%
where $\epsilon_l=(2l+1)\pi T$ is a Fermi Matsubara frequency,
$\omega_m=2m\pi T$ is a Bose Matsubara frequency, and $a$, $L$ are the
width and length of the wire respectively.  The diffusive form of the
denominators in Eqn.~(\ref{bubble}) is valid only for momentum and
frequency values such that $Dq^{\prime 2},\ |\epsilon_l|,\
|\omega_m|<1/\tau$, so we introduce the upper cutoff, $1/\tau$, in all
these sums.

For the rest of this section we will work at zero magnetic field, so
we set $q=0$.  The terms $\delta P_{1-5}$ have previously been summed
for 2D films\cite{SRW95} to give the expression
%%%%%%%%%%%%%%%%%%%%%%%%%%%%%%%%%%%%%%%%%%%%%%%%%%%%%%%%%%%%%%%%%%%%%%%%%%%%%%%%%%%%%%%%
\begin{eqnarray}
\label{fullpert}
\ln \left( \frac{T_c}{T_{c0}} \right) &=&
-T\sum_{\omega_m}{1\over aL}\sum_{q^{\prime}}\left\{
\displaystyle\phantom{\frac{\displaystyle\frac{1}{2}^2}{\displaystyle\frac{1}{2}^2}}
\!\!\!\!\!\!-{1\over 2\pi T}{Dq^{\prime 2}\over\omega_m^2-(Dq^{\prime 2})^2}
\psi'\left({1\over 2}+{\omega_m\over 2\pi T}\right)
+{2Dq^{\prime 2} [\omega_m^2+(Dq^{\prime
2})^2]\over\omega_m[\omega_m^2-(Dq^{\prime 2})^2]^2}
\left[\psi\left({1\over 2}+{\omega_m\over 2\pi T}\right)
-\psi\left({1\over 2}\right)\right]\right.\nonumber\\
&-&\left.{\displaystyle 4(Dq^{\prime 2})^2\over [\omega_m^2-(Dq^{\prime 2})^2]^2}
{\displaystyle\left[\psi\left({1\over 2}+{\omega_m\over 2\pi T}\right)
-\psi\left({1\over 2}\right)\right]^2\over
\displaystyle\left[\psi\left({1\over 2}+{Dq^{\prime 2}+\omega_m\over 4\pi T}\right)
-\psi\left({1\over 2}\right)\right]}\right\}
V_C(q^{\prime},\omega_m)
\end{eqnarray}
%%%%%%%%%%%%%%%%%%%%%%%%%%%%%%%%%%%%%%%%%%%%%%%%%%%%%%%%%%%%%%%%%%%%%%%%%%%%%%%%%%%%%%%%
The $1/q^{\prime 2}$ singularity in the screened Coulomb potential
$V_C(q^{\prime},\omega_m)$ is cancelled since there is an overall
factor of $Dq^{\prime 2}$ in the above equation.  Note that there is,
in fact, no divergence at $Dq^{\prime 2}=\omega_m$, but this can only
be seen by Taylor expanding to second order. The need to perform this
expansion makes the calculation involved, especially at the level of
numerics. To allow simplification, we define $y=Dq^{\prime 2}/2\pi T$
and $m=\omega_m/2\pi T$, which leads to the result
%%%%%%%%%%%%%%%%%%%%%%%%%%%%%%%%%%%%%%%%%%%%%%%%%%%%%%%%%%%%%%%%%%%%%%%%%%%%%%%%%%%%%%%%%
\begin{eqnarray}
\label{fpertlong}
\ln \left( \frac{T_c}{T_{c0}} \right) &=&
-{1\over 4\pi^2N(0)T}\sum_{m=1}^M{1\over aL}\sum_{q^{\prime}}\left[
\displaystyle\phantom{\frac{\displaystyle\frac{1}{2}^2}{\displaystyle\frac{1}{2}^2}}\!\!\!\!\!\!
\left\{
{2\over m}\left[\psi\left({1\over 2}+m\right)-\psi\left({1\over 2}\right)\right]
+\psi'\left({1\over 2}+m\right)\right\}{1\over y+m}\right.\nonumber\\
&-&{4y\over (m+y)(m-y)^2}\left.\left\{
{\displaystyle\left[\psi\left({1\over 2}+m\right)-\psi\left({1\over 2}\right)\right]^2\over
\displaystyle\left[\psi\left({1\over 2}+{m+y\over 2}\right)-\psi\left({1\over 2}\right)\right]}
-\left[\psi\left({1\over 2}+m\right)-\psi\left({1\over 2}\right)\right]
+{1\over 2}(y-m)\psi'\left({1\over 2}+m\right)\right\}\right]
\end{eqnarray}
%%%%%%%%%%%%%%%%%%%%%%%%%%%%%%%%%%%%%%%%%%%%%%%%%%%%%%%%%%%%%%%%%%%%%%%%%%%%%%%%%%%%%%%%%%
where we have split off the most divergent term at large momentum and
frequency, and $M=1/2\pi T\tau$.  To proceed further we must consider
the $q^{\prime}$-sum in various dimensions. In two dimensions the sum
becomes an integral, and it is most sensible to perform the cut-off in
a circularly symmetric manner
%%%%%%%%%%%%%%%%%%%%%%%%%%%%%%%%%%%%%%%%%%%%%%%%%%%%%%%%%%%%%%%%%%%%%%%%%%%%%%%%%%%%%%%%%%
\begin{equation}
\label{2Dsum}
{1\over aL}\sum_{q^{\prime}} \mapsto \int_{0}^{1/\tau} {2\pi
q^{\prime} dq^{\prime}\over (2\pi)^2} ={T\over 2D} \int_{0}^{M} dy.
\end{equation}
%%%%%%%%%%%%%%%%%%%%%%%%%%%%%%%%%%%%%%%%%%%%%%%%%%%%%%%%%%%%%%%%%%%%%%%%%%%%%%%%%%%%%%%%%%
In quasi-1D wires, we must replace the integral in the transverse direction by a sum,
%%%%%%%%%%%%%%%%%%%%%%%%%%%%%%%%%%%%%%%%%%%%%%%%%%%%%%%%%%%%%%%%%%%%%%%%%%%%%%%%%%%%%%%%%%
\begin{equation}
\label{1Dsum}
{1\over aL}\sum_{q^{\prime}} \mapsto {1\over a}\sum_{q_a'}\int_{0}^{1/\tau} {dq_L'\over 2\pi},
\end{equation}
%%%%%%%%%%%%%%%%%%%%%%%%%%%%%%%%%%%%%%%%%%%%%%%%%%%%%%%%%%%%%%%%%%%%%%%%%%%%%%%%%%%%%%%%%%
where $q^{\prime 2}=q_L'^2+q_a'^2$.  At this point we have to decide
which boundary conditions to use in the transverse direction.  To
illustrate the strong effect this has we will compare periodic and
zero-gradient boundary conditions, the details of which are listed in
Table 1.  The zero-gradient boundary conditions are the physically
correct ones for a superconductor-insulator boundary as discussed in
Ref. [\cite{deG}].  To perform the sum over $q^{\prime}$ note that we
can write $y=y_L+y_a$ where $y_L=Dq_L'^2/2\pi T$ and $y_a=Dq_a'^2/2\pi
T$, and if we define $z=\sqrt{y_L}$, the $q_L'$-integral becomes
%%%%%%%%%%%%%%%%%%%%%%%%%%%%%%%%%%%%%%%%%%%%%%%%%%%%%%%%%%%%%%%%%%%%%%%%%%%%%%%%%%%%%%%%%%
\begin{equation}
\label{1Dint}
\int_{0}^{1/\tau} {dq^{\prime}_L\over 2\pi} = \int_{0}^{\sqrt{M}} {dz\over L_T}.
\end{equation}
%%%%%%%%%%%%%%%%%%%%%%%%%%%%%%%%%%%%%%%%%%%%%%%%%%%%%%%%%%%%%%%%%%%%%%%%%%%%%%%%%%%%%%%%%%
where $L_T=\sqrt{2\pi D/T}$ is the thermal length. Note that
$y=z^2+4p(a/L_T)^2$ for zero-gradient boundary conditions and
$y=z^2+p(a/L_T)^2$ for periodic boundary conditions.  The final result
of perturbation theory is then given by Eqn. (\ref{fpertlong}) with
the prefactor and sums becoming
%%%%%%%%%%%%%%%%%%%%%%%%%%%%%%%%%%%%%%%%%%%%%%%%%%%%%%%%%%%%%%%%%%%%%%%%%%%%%%%%%%%%%%%%%%
\begin{equation}
\label{prefactor}
-{1\over 4\pi^2N(0)T}\sum_{m=1}^{M}{1\over
aL}\sum_{q^{\prime}}\quad\mapsto\quad
\cases{-\displaystyle t\sum_{m=1}^{M}\int_{0}^{M} dy&\quad\hbox{2D}\cr
-\displaystyle{t\over\pi}\left({L_T\over a}\right)\sum_{m=1}^M\sum_p\int_{0}^{\sqrt{M}}dz&
\quad\hbox{quasi--1D}\cr}
\end{equation}
%%%%%%%%%%%%%%%%%%%%%%%%%%%%%%%%%%%%%%%%%%%%%%%%%%%%%%%%%%%%%%%%%%%%%%%%%%%%%%%%%%%%%%%%%%% 
In the above, $t$ is a dimensionless measure of disorder given by
%%%%%%%%%%%%%%%%%%%%%%%%%%%%%%%%%%%%%%%%%%%%%%%%%%%%%%%%%%%%%%%%%%%%%%%%%%%%%%%%%%%%%%%%
\begin{equation}
t= \frac{1}{8 \pi^2 N(0)D}=\frac{R_{\square}}{R_0}
\end{equation}
%%%%%%%%%%%%%%%%%%%%%%%%%%%%%%%%%%%%%%%%%%%%%%%%%%%%%%%%%%%%%%%%%%%%%%%%%%%%%%%%%%%%%%%%
where $R_{\square}$ is the normal state resistance per square, and
$R_0=2\pi h / e^2 \approx 162 k \Omega$. Note that $t$ only differs
from the previously discussed parameter $1/k_F\ell$ by a factor
$2\pi$.

Having shown that the low-momentum singularity in the screened Coulomb
potential is cancelled by diagram 5, we recalculate the suppression of
transition temperature using a featureless Coulomb potential of
magnitude $N(0)V_C=1/2$, and including only diagrams 1--4. We have
previously shown\cite{SHA00} that the above equations simplify to give
%%%%%%%%%%%%%%%%%%%%%%%%%%%%%%%%%%%%%%%%%%%%%%%%%%%%%%%%%%%%%%%%%%%%%%%%%%%%%%%%%
\begin{eqnarray}
\label{P1234}
\delta P_{12}&=&-T\sum_{\omega_m>0}\left[
\psi\left({1\over 2}+{\omega_m\over 2\pi T}\right)
-\psi\left({1\over 2}\right)\right]\sum_{q^{\prime}} {1\over
(Dq^{\prime 2}+\omega_m)^2}\nonumber\\
\delta P_{34}&=&-2T\sum_{\omega_m>0}\left[
\psi\left({1\over 2}+{\omega_m\over 2\pi T}\right)
-\psi\left({1\over 2}\right)\right]{1\over\omega_m}\sum_{q^{\prime}}
{1\over (Dq^{\prime 2}+\omega_m)}.
\end{eqnarray}
%%%%%%%%%%%%%%%%%%%%%%%%%%%%%%%%%%%%%%%%%%%%%%%%%%%%%%%%%%%%%%%%%%%%%%%%%%%%%%%%%
For convenience we define dimensionless sums $I_1(m)$ and $I_2(m)$ as
%%%%%%%%%%%%%%%%%%%%%%%%%%%%%%%%%%%%%%%%%%%%%%%%%%%%%%%%%%%%%%%%%%%%%%%%%%%%%%%%
\begin{eqnarray}
\label{Idef}
I_1(m)&\equiv& {4\pi D\over aL}\sum_{q^{\prime}} \frac{1}{Dq^{\prime 2} +|\omega_m|}\nonumber\\
I_2(m)&\equiv& {4\pi D\over aL}\sum_{q^{\prime}} \frac{2\pi T}{(Dq^{\prime 2} +|\omega_m|)^2}
\end{eqnarray}
%%%%%%%%%%%%%%%%%%%%%%%%%%%%%%%%%%%%%%%%%%%%%%%%%%%%%%%%%%%%%%%%%%%%%%%%%%%%%%%%
which are evaluated in two dimensions and quasi-one dimension in
Appendix A.  When we combine Eqns. (\ref{Tcdef}) and (\ref{P1234}) we
obtain the formula for transition temperature suppression assuming a
featureless Coulomb potential,
%%%%%%%%%%%%%%%%%%%%%%%%%%%%%%%%%%%%%%%%%%%%%%%%%%%%%%%%%%%%%%%%%%%%%%%%%%%%%%%%%%%%%%%
\begin{equation}
\label{pert1234}
\ln \left( \frac{T_c}{T_{c0}} \right) =
\displaystyle -t\sum_{m=1}^{M} \left[\psi\left({1\over 2}+m\right)
-\psi\left({1\over 2}\right)\right]
\left[  I_2(m)+\frac{2I_1(m)}{m} \right].
\end{equation}
%%%%%%%%%%%%%%%%%%%%%%%%%%%%%%%%%%%%%%%%%%%%%%%%%%%%%%%%%%%%%%%%%%%%%%%%%%%%%%%%%%%%%%%%
If we ignore the self-energy diagrams ($\delta P_{12}$), the $I_2(m)$ term in the above
equation vanishes to leave
%%%%%%%%%%%%%%%%%%%%%%%%%%%%%%%%%%%%%%%%%%%%%%%%%%%%%%%%%%%%%%%%%%%%%%%%%%%%%%%%%%%%%%%
\begin{equation}
\label{pert34}
\ln \left( \frac{T_c}{T_{c0}} \right) =
\displaystyle -2t\sum_{m=1}^{M} \left[\psi\left({1\over 2}+m\right)
-\psi\left({1\over 2}\right)\right]
\frac{2I_1(m)}{m}.
\end{equation}
%%%%%%%%%%%%%%%%%%%%%%%%%%%%%%%%%%%%%%%%%%%%%%%%%%%%%%%%%%%%%%%%%%%%%%%%%%%%%%%%%%%%%%%%

Having derived expressions for the suppression of transition
temperature using various approximations and boundary conditions, we
perform a detailed numerical analysis, the results of which can be
summarised as follows. The transition temperature is very sensitive to
the boundary conditions used -- periodic boundary conditions do not
yield a dependence of transition temperature on wire width, whilst
zero-gradient boundary conditions show a systematic suppression as
wire width is decreased.  The predictions are also sensitive to the
value of the thermal length, $L_T$, since the width, $a$, enters only
in the dimensionless combination $a/L_T$. Experimentally $L_T$ may be
known only roughly and it may be necessary to fit its value. On the
other hand, the predictions are not very sensitive to either the form
of the Coulomb interaction used (Eqns. (\ref{fpertlong}) and
(\ref{pert1234})), or the inclusion of self-energy corrections
(Eqns. (\ref{pert1234}) and (\ref{pert34})).  The slight difference
between these approaches vanishes when we fit the initial slope of the
$T_c(R_{\square})$ curve to experiment, which sets the value of the
BCS attraction, $\gamma$. This demonstrates that the simplest approach
-- featureless Coulomb potential and no self-energy diagrams -- is
valid. This is important since these simplifications are necessary for
the resummation technique discussed in the next section. It is
necessary to use this technique to cure the unphysical re-entrance
problems which beset perturbation theory at low temperature.

\section{The Resummation Technique}

In this section we will review the resummation technique\cite{OF99}
developed by Oreg and Finkel'stein (OF) as applied to the problem of
$T_c$ suppression in quasi-1D wires. This method involves calculating
the pair-amplitude, $\Gamma(\epsilon_n,\epsilon_l)$, which is a matrix
with Fermi Matsubara frequencies labelling its rows and columns.  The
transition temperature is identified as the temperature at which this
matrix becomes singular. OF obtained an equation for $\Gamma$ by
including only pseudopotential diagrams, and solved it using matrix
perturbation theory with zero-gradient boundary
conditions\cite{OFBC}. We have performed calculations to show that it
is legitimate to include only these diagrams, and that it is essential
to use zero-gradient boundary conditions, as described in section II
for perturbation theory. In a previous paper\cite{SHA00} we have shown
that the equation for $\Gamma$ can be solved exactly at no extra
numerical cost, so the matrix perturbation theory is unnecessary, and
indeed gives incorrect results at large disorder.  Therefore we shall
include only pseudopotential diagrams and solve the equation for
$\Gamma$ using an exact diagonalization method with zero-gradient
boundary conditions.

Before we proceed further, let us briefly describe the details of the technique.
The ladder summation involved is shown in Fig. 2. The block $t\Lambda$, 
an effective Coulomb pseudopotential, contains diagrams 3 and 4 of Fig. 1(c);
the block $\Sigma$, the self-energy of
the Cooperon impurity ladder, $C$, contains diagrams 1 and 2 of Fig. 1(c).
We have included the self-energy diagrams for
completeness, although their effects can be ignored in calculations, as discussed above.
The equation for $\Gamma$ is seen to be
%%%%%%%%%%%%%%%%%%%%%%%%%%%%%%%%%%%%%%%%%%%%%%%%%%%%%%%%%%%%%%%%%%%%%%%%%%%%%%%%%%%%
\begin{equation}
\label{OFeqn}
\Gamma(\epsilon_n,\epsilon_l)=-|\gamma|+t\Lambda(\epsilon_n,\epsilon_l)
-2\pi T\sum_{m=0}^M
\left[-|\gamma|+t\Lambda(\epsilon_n,\epsilon_m)\right] {1\over
|\epsilon_m|}\Gamma(\epsilon_m,\epsilon_l)
\end{equation}
%%%%%%%%%%%%%%%%%%%%%%%%%%%%%%%%%%%%%%%%%%%%%%%%%%%%%%%%%%%%%%%%%%%%%%%%%%%%%%%%%%%%
where the contributions to $t\Lambda$ are
%%%%%%%%%%%%%%%%%%%%%%%%%%%%%%%%%%%%%%
\begin{eqnarray}
t\Lambda_3(\epsilon_n,\epsilon_m)&=&\displaystyle {1\over 2\pi
N(0)}{1\over aL}\sum_{q^{\prime}} {Dq^2+Dq^{\prime
2}+|\epsilon_n|+|\epsilon_m|\over [Dq^{\prime
2}+|\epsilon_n|+|\epsilon_m|]^2}\theta(-\epsilon_n\epsilon_m)\nonumber\\
\displaystyle t\Lambda_4(\epsilon_n,\epsilon_m) &=&\displaystyle
{1\over 2\pi N(0)}{1\over aL}\sum_{q^{\prime}} {1\over [Dq^{\prime
2}+|\epsilon_n|+|\epsilon_m|]}
\theta(\epsilon_n\epsilon_m).
\label{lambda}
\end{eqnarray}
%%%%%%%%%%%%%%%%%%%%%%%%%%%%%%%%%%%%%%%
We have kept non-zero external momentum $q$ so that we can include
magnetic field effects later. For now we consider the case of zero
magnetic field, so we set $q=0$, and the two contributions to
$t\Lambda$ are identical.  It follows that
$\Lambda(\epsilon_n,\epsilon_m)$ does not depend upon the relative
sign of $\epsilon_n$ and $\epsilon_m$ and has the form
%%%%%%%%%%%%%%%%%%%%%%%%%%%%%%%%%%%%%%%%
\begin{equation}
\Lambda(\epsilon_n,\epsilon_m)=
I_1(n+m+1),
\label{lambdaq=0}
\end{equation}
%%%%%%%%%%%%%%%%%%%%%%%%%%%%%%%%%%%%%%%%
where $I_1$ is defined in Eqn. (\ref{Idef}).
Eqn. (\ref{OFeqn}) can then be written in matrix form as
%%%%%%%%%%%%%%%%%%%%%%%%%%%%%%%%%%%%%
\begin{equation}
{\hat\Gamma}=-|\gamma|{\hat 1}+t{\hat\Lambda}
-[-|\gamma|{\hat 1}+t{\hat\Lambda}]{\hat\epsilon}^{-1}{\hat\Gamma},
\end{equation}
%%%%%%%%%%%%%%%%%%%%%%%%%%%%%%%%%%%%%
where
${\hat\Gamma}_{nm}=\Gamma(\epsilon_n,\epsilon_m)$, ${\hat 1}_{nm}=1$,
${\hat\Lambda}_{nm}=\Lambda(\epsilon_n,\epsilon_m)$ and
${\hat\epsilon}_{nm}=(n+1/2)\delta_{nm}$. This has the solution
%%%%%%%%%%%%%%%%%%%%%%%%%%%%%%%%%%%%%
\begin{equation}
{\hat\Gamma}={\hat\epsilon}^{1/2}({\hat I}-|\gamma|{\hat\Pi})^{-1}
{\hat\epsilon}^{-1/2}(-|\gamma|{\hat 1}+t{\hat\Lambda}),
\end{equation}
%%%%%%%%%%%%%%%%%%%%%%%%%%%%%%%%%%%%%
where
%%%%%%%%%%%%%%%%%%%%%%%%%%%%%%%%%%%%%
\begin{equation}
{\hat\Pi}={\hat\epsilon}^{-1/2}
[{\hat 1}-|\gamma|^{-1}t{\hat\Lambda}]{\hat\epsilon}^{-1/2},
\end{equation}
%%%%%%%%%%%%%%%%%%%%%%%%%%%%%%%%%%%%%
and ${\hat I}_{nm}=\delta_{nm}$ is the identity matrix. It follows that
when the matrix ${\hat\Pi}$ has an eigenvalue equal to $1/|\gamma|$, the
pair amplitude diverges, and we have found $T_c$. In other words, we need
to solve the eigenvalue equation
%%%%%%%%%%%%%%%%%%%%%%%%%%%%%%%%%%%%%%%%%%%%%%%%%%%%%%%%%%%%%%%%%%%%%%%%%%%%
\begin{equation}
\label{eigeqn}
\left[|\gamma|^{-1}{\hat I}-{\hat\Pi}(T_c)\right]
\left|\psi\right.\rangle=0.
\end{equation}
%%%%%%%%%%%%%%%%%%%%%%%%%%%%%%%%%%%%%%%%%%%%%%%%%%%%%%%%%%%%%%%%%%%%%%%%%%% 

We solve Eqn. (\ref{eigeqn}) by diagonalizing the matrix
$[|\gamma|^{-1}{\hat I}-{\hat\Pi}(T)]$ numerically, and decreasing the
temperature until its lowest eigenvalue equals zero. Since we only
need the lowest eigenvalue, and the matrix has such a simple
structure, the Lanczos method is very efficient here.  The only minor
complication is that the matrix ${\hat\Pi}(T)$ depends upon $T$ both
through the dependence of $\hat\Lambda$ upon $T$ and through its rank
$M=1/2\pi T\tau$. We therefore start at the mean-field value of $M$,
which we will call $M_0$, and decrease the temperature by increasing
$M$ successively by one.  We diagonalize the matrix
$[|\gamma|^{-1}{\hat I}-{\hat\Pi}]$ for each value of $M$ until the
lowest eigenvalue changes sign. At this point we have found $T_c$ for
the given problem, and $T_c/T_{c0}=M_0/M$. This is obviously the
method of choice, since no approximation is involved.

We can now compare theory and experiment: there are three sets of
experimental data that we know about -- those of Graybeal et al
(1987)\cite{GMDB87}, Sharifi et al (1993)\cite{SHD93}, and Xiong et al
(1997)\cite{XHD97}. The data of Graybeal et al consists of $T_c$ vs
$a$ for wires of constant thickness (and hence constant resistance per
square). Their data shows that the relative suppression of $T_c$
varies as $1/a^2$ with a proportionality constant independent of
$R_{\square}$. We have performed calculations which predict a
dependence roughly of the form $1/a$, but with a proportionality
constant dependent on $R_{\square}$.  We will therefore not consider
this data further. The data of Sharifi et al and Xiong et al are both
of the form of $T_c(R_{\square},a)$ vs $R_{\square}$ for wires of
different width, but only the latter shows a consistent trend with
decreasing width, so it is only this data that we shall attempt to
fit.  The fact that this is the most recent data is reassuring, since
it suggests that more recent fabrication techniques have produced
better quality wires.

In Fig. 3 we compare the data of Xiong et al with our theoretical
predictions. In doing this we encounter two difficulties. The first is
that experiment and theory do not agree for some of the 2D films
used. We assume that this is due to some additional physical processes
that will affect samples in the same manner independent of width. We
therefore multiply our theoretical predictions by
$T_c(\hbox{2D,expt})/T_c(\hbox{2D,theory})$ to factor out this
effect. The second difficulty is that wires of different widths do not
appear to have the same value of $T_c$ in the clean limit,
$T_c(0,a)$. We therefore assume a different $T_c(0,a)$ for each
wire. Upon making these two assumptions, we see that the agreement
between theory and experiment is good. However plotting the data in
this form emphasises the suppression of transition temperature due to
disorder, whereas we are interested in the additional suppression due
to the decrease in width. For this reason we need to replot the data
to emphasise this latter point.

In Fig. 4 we replot Fig. 3 in terms of the {\it additional relative}
suppression, $\Theta(R_{\square},a)$, which is defined by
%%%%%%%%%%%%%%%%%%%%%%%%%%%%%%%%%%%%%%%%%%%%%%%%%%
\begin{equation}
\Theta(R_{\square},a)=
\frac{T_c(R_{\square},a)/T_c(0,a)}{T_c(R_{\square},2D)/T_c(0,2D)}.
\end{equation}
%%%%%%%%%%%%%%%%%%%%%%%%%%%%%%%%%%%%%%%%%%%%%%%%%%
This quantity focuses on the basic physical phenomenon we are
interested in -- the additional suppression of transition temperature
caused by the reduced dimensionality in the quasi-1D wires. It has the
additional benefit of factoring out the two difficulties discussed in
the previous paragraph. When plotted in this form, the agreement
between theory and experiment no longer appears to be as good. This is
because our theory, and indeed any theory, predicts a smooth
dimensional crossover whereas the experimental data shows an abrupt
crossover between the 2D film and the $1000\AA$ wire, which lie on one
curve, and the $580\AA, 350\AA, 250\AA$ wires, which lie on a
different curve. The fact that the data for the three narrowest wires
fall on top of each other might be due to them having the same
effective dimensionless width $a/L_T$. If we allowed ourselves the
additional freedom of fitting this parameter separately for each wire,
we would obtain good agreement between theory and experiment, but we
do not believe this is justified. We have used the value of the
thermal length quoted in Sharifi et al (viz $L_T=400\AA$ at $T=4K$,
which corresponds to $L_T=300\AA$ at $T_{c0}=6.9K$).

The basic problem is clearly that there is not enough experimental
data available -- there are only 19 points in total, and each of these
corresponds to a different wire. To carry out a more systematic
comparison between theory and experiment, one would need wires of many
different resistances (thicknesses) for each value of width. This
would, however, be very costly in terms of experimental effort. A less
costly method for generating more data would be to apply a magnetic
field to a given wire.  We therefore extend our theory to include the
effect of such a field in the next section.

\section{Effect of an Applied Magnetic Field}

We will now consider the effect of an applied perpendicular magnetic
field.  There are two effects to consider here: the {\it orbital}
effect, caused by the magnetic field scrambling the relative phase of
the electrons in a Cooper pair, and the {\it Zeeman} effect, caused by
the magnetic field lifting the degeneracy of up and down spins. Let us
first consider the relative importance of the two effects. The orbital
effect will destroy superconductivity when the magnetic dephasing
rate, $1/\tau_H$, is of the order of $k_BT_c$, which gives $H\approx
4T$ for a Pb wire of width $250\AA$.  The Zeeman splitting will
destroy superconductivity when its magnitude $\mu_BH$ is of the order
of $k_BT_c$, which corresponds to $H\approx 10T$ for Pb. Since these
fields have the same order of magnitude, we will need to consider both
effects.

We first examine the orbital effect. The magnetic field causes the eigenvalues
of the Cooperon operator to change from the zero-field values
%%%%%%%%%%%%%%%%%%%%%%%%%%%%%%%%%%%%%%%%%%%%%%%%%%%%%%%%%%%%%%%%
\begin{equation}
Dq^2=D\left[\left({2\pi n\over L}\right)^2+
\left({\pi p\over a}\right)^2\right],
\qquad n=0,\pm 1,\pm 2\dots, \quad p=0,1,2\dots
\end{equation}
%%%%%%%%%%%%%%%%%%%%%%%%%%%%%%%%%%%%%%%%%%%%%%%%%%%%%%%%%%%%%%%%
to the eigenvalues, $\lambda$, of the minimally coupled Cooperon operator
%%%%%%%%%%%%%%%%%%%%%%%%%%%%%%%%%%%%%%%%%%%%%%%%%%%%%%%%%%%%%%%%
\begin{equation}
D(-i{\bf\nabla}+2e{\bf A})^2\psi(x,y) 
=D\left[\left(-i{\partial\over\partial x}+2eHy\right)^2
+\left(-i{\partial\over\partial y}\right)^2\right]\psi(x,y)
=\lambda\psi(x,y),
\end{equation}
%%%%%%%%%%%%%%%%%%%%%%%%%%%%%%%%%%%%%%%%%%%%%%%%%%%%%%%%%%%%%%%%
where we have chosen the gauge ${\bf A}=(-Hy,0,0)$.  The
eigenfunctions $\psi(x,y)$ must satisfy zero-gradient boundary
conditions on the sides of the wire, $y=\pm a/2$, and periodic
boundary conditions at the ends of the wire, $x=0,L$. The solution has
the separable form
%%%%%%%%%%%%%%%%%%%%%%%%%%%%%%%%%%%%%%%%%%%%%%%%%%%%%%%%%%%%%%%%
\begin{equation}
\psi(x,y)=\phi_{q_n,p}(y)e^{iqx},\qquad
q_n={2\pi n\over L},\quad
n=0,\pm 1,\pm 2\dots
\end{equation}
%%%%%%%%%%%%%%%%%%%%%%%%%%%%%%%%%%%%%%%%%%%%%%%%%%%%%%%%%%%%%%%%
where $\phi_{q_n,p}$ satisfies the equation
%%%%%%%%%%%%%%%%%%%%%%%%%%%%%%%%%%%%%%%%%%%%%%%%%%%%%%%%%%%%%%%%
\begin{equation}
D\left[-{d^2\over dy^2}+(q+2eHy)^2\right]\phi_{q_n,p}(y)
=\lambda_{q_n,p}\phi_{q_n,p}(y),\qquad
{d\phi_{q_n,p}\over dy}(\pm a/2)=0,
\end{equation}
%%%%%%%%%%%%%%%%%%%%%%%%%%%%%%%%%%%%%%%%%%%%%%%%%%%%%%%%%%%%%%%%
and $p=0,1,2\dots$ labels the eigenvalues in ascending order.

To solve this eigenvalue problem numerically, we convert it into a
matrix diagonalization problem by working in a truncated function
space spanned by eigenfunctions of the zero-field problem.  The lowest
eigenvalue, $\lambda_{0,0}$, equals $2/\tau_H$, where $1/\tau_H$ is
the magnetic dephasing rate which appears in the Abrikosov-Gor'kov
equation\cite{AG61} for $T_c(H)$.  There are two asymptotic limits
where $1/\tau_H$ is easily determined: the 2D limit, where the problem
becomes that of a simple harmonic oscillator, with $1/\tau_H=DeH$; and
the 1D limit, where the eigenfunction is simply a constant, and hence
(using perturbation theory) $1/\tau_H=De^2H^2a^2/6$. The 2D to 1D
crossover occurs when the width, $a$, is of the order of the magnetic
length, $L_H=(\hbar/eH)^{1/2}$, which occurs at the field
$H_0=h/ea^2$. At larger magnetic fields the cyclotron orbit of the
electron lies entirely within the wire, and hence we are in the 2D
limit; at smaller magnetic fields the electron hits the sides of the
wire before completing a cyclotron orbit, and hence we are in the 1D
limit.  To summarise,
%%%%%%%%%%%%%%%%%%%%%%%%%%%%%%%%%%%%%%%%%%%%%%%%%%%%%%%%%%%%%%%%
\begin{equation}
{1\over\tau_H}=\cases{
DeH\quad&$H\gg H_0$ ($a\gg L_H$)\quad 2D\cr
De^2H^2a^2/6\quad&$H\ll H_0$ ($a\ll L_H$)\quad 1D\cr}
\end{equation}
%%%%%%%%%%%%%%%%%%%%%%%%%%%%%%%%%%%%%%%%%%%%%%%%%%%%%%%%%%%%%%%%

These eigenvalues then enter into our equations for the pair
propagator in two ways: (i) the term $Dq^2$ in the $q$-dependent pair
propagator is replaced by the lowest eigenvalue $\lambda_{0,0}$; (ii)
sums over functions of $Dq^{\prime 2}$ in diagrams which involve
Cooperonic ladders (e.g. diagrams 2 and 4 in Fig. 1(c)) are replaced
by sums over $\lambda_{q_n,p}$.  Obviously sums over functions of
$Dq^{\prime 2}$ in diagrams which involve diffusion ladders
(e.g. diagrams 1 and 3 in Fig. 1(c)) remain unchanged.  It is
numerically very expensive to perform the sums over all the
eigenvalues, so we would like to see how important it is to use these
eigenvalues in any sums.  To answer this question we again use
perturbation theory as our testing ground.  We perform the Cooperonic
sum in two different ways -- both correctly, and by replacing it by
the corresponding diffusion sum. In both cases we still make the
replacement of $Dq^2$ by $\lambda_{0,0}=2/\tau_H$. The perturbative
equations are then
%%%%%%%%%%%%%%%%%%%%%%%%%%%%%%%%%%%%%%%%%
\begin{eqnarray}
{\delta P_1\over N(0)}&=&-{t\over 2}\sum_{m=1}^{M} \left\{
2\left[\psi\left({1\over 2}+m+\alpha\right)-
\psi\left({1\over 2}+\alpha\right)\right]I_2(m)+
\left[\psi'\left({1\over 2}+\alpha\right)-
\psi'\left({1\over 2}+m+\alpha\right)\right]I_1(m)\right\}\nonumber\\
{\delta P_2\over N(0)}&=&+{t\over 2}\sum_{m=1}^{M}
\left[\psi'\left({1\over 2}+\alpha\right)-
\psi'\left({1\over 2}+m+\alpha\right)\right]C_1(m)\nonumber\\
{\delta P_3\over N(0)}&=&-t\sum_{m=1}^{M} {1\over m+2\alpha}
\left[\psi\left({1\over 2}+m+\alpha\right)-
\psi\left({1\over 2}+\alpha\right)\right](I_1(m)+2\alpha I_2(m) )\nonumber\\
{\delta P_4\over N(0)}&=&-t\sum_{m=1}^{M} {1\over m+2\alpha}
\left[\psi\left({1\over 2}+m+\alpha\right)-
\psi\left({1\over 2}+\alpha\right)\right]C_1(m)
\end{eqnarray}
%%%%%%%%%%%%%%%%%%%%%%%%%%%%%%%%%%%%%%%%%
where $\alpha=1/2\pi T\tau_H$. The diffusive sums $I_1(m)$ and
$I_2(m)$ are defined as in Eqn. (\ref{Idef}) except that here we
perform the momentum sums in the longitudinal direction {\it exactly}
to give
%%%%%%%%%%%%%%%%%%%%%%%%%%%%%%%%%%%%%%%%%
\begin{eqnarray}
I_1(m)&=&{4\pi D\over aL}\sum_{q} {1\over Dq^2+2\pi Tm} ={L_T^2\over
\pi La}\sum_{n=-n_0}^{n_0}\sum_{k=0}^{k_0}
{1\over\left[m+\left({kL_T\over 2a}\right)^2+\left({nL_T\over
L}\right)^2\right]}
\nonumber\\
I_2(m)&=&{4\pi D\over aL}\sum_{q} {2\pi T\over (Dq^2+2\pi Tm)^{2}}
={L_T^2\over \pi La}\sum_{n=-n_0}^{n_0}\sum_{k=0}^{k_0}
{1\over\left[m+\left({kL_T\over 2a}\right)^2+\left({nL_T\over
L}\right)^2
\right]^{2}}
\end{eqnarray}
%%%%%%%%%%%%%%%%%%%%%%%%%%%%%%%%%%%%%%%%%
and the Cooperonic sum, $C_1$, which is now distinct from $I_1$, is
given by
%%%%%%%%%%%%%%%%%%%%%%%%%%%%%%%%%%%%%%%%%
\begin{equation}
C_{1}(m)={4\pi D\over aL}\sum_{q} {1\over Dq^2+2\pi Tm}= {L_T^2\over
\pi La}\sum_{n=-n_0}^{n_0}\sum_{k=0}^{k_0} {1\over
\left[m+\lambda_{nk}/2\pi T\right]}.
\end{equation}
%%%%%%%%%%%%%%%%%%%%%%%%%%%%%%%%%%%%%%%%%
We found essentially no difference between using the correct $C_1$, or
replacing it by $I_1$, over a wide range of magnetic fields. It
follows that the only replacement we need to make is of $Dq^2$ by
$\lambda_{0,0}$, which greatly reduces the numerical cost.

Let us next consider the effect of the Zeeman splitting. This leads to
changes in both the mean field and perturbative terms, and we will
show that only the former effect is important. To do this we will
evaluate the mean-field pair polarization function and the first-order
perturbation correction, $\delta P$, both with and without Zeeman
splitting. We find that the latter is essentially unaffected.  To
treat the Zeeman effect, we need formulas for the Cooperon impurity
ladder and Hikami boxes\cite{Hik81}, where we note that the boxes in
diagrams 1 and 3 of Fig. 1(c) will give different contributions. The
non-zero elements of the spin-dependent Cooperon ladder,
$C_{\alpha\beta\gamma\delta}$, are
%%%%%%%%%%%%%%%%%%%%%%%%%%%%%%%%%%%%%%%%%
\begin{equation}
C_{\uparrow\uparrow\downarrow \downarrow}={1\over 2\pi N(0)\tau^2}
{1\over Dq^2+|\omega|+|\omega'|-2i\Delta},\quad
C_{\downarrow\downarrow\uparrow\uparrow}={1\over 2\pi N(0)\tau^2}
{1\over Dq^2+|\omega|+|\omega'|+2i\Delta}=
C^*_{\uparrow\uparrow\downarrow\downarrow}
\end{equation}
%%%%%%%%%%%%%%%%%%%%%%%%%%%%%%%%%%%%%%%%%
where $\Delta=\mu_B H$ is the Zeeman splitting. The appropriate spin
combination which occurs in an s-wave pair propagator is
$C_{\uparrow\uparrow\downarrow
\downarrow}+C_{\downarrow\downarrow\uparrow\uparrow} =2\Re e
C_{\uparrow\uparrow\downarrow\downarrow}$. This leads to the mean
field formula for $T_c$,
%%%%%%%%%%%%%%%%%%%%%%%%%%%%%%%%%%%%%%%%%
\begin{equation}
\ln{\left({T_c\over T_{c0}}\right)}=\psi\left({1\over 2}\right)
-\Re e \left[\psi\left({1\over 2}+\kappa\right)\right]
\end{equation}
%%%%%%%%%%%%%%%%%%%%%%%%%%%%%%%%%%%%%%%%%
where $\kappa=\alpha-i\beta$ with $\beta=\Delta/2\pi T_c=\mu_B H/2\pi
T_c$.  To calculate the first-order pair polarization terms $\delta
P_{1-4}$, we note that: (a) the diffusion ladders on Coulomb vertices
are unchanged; (b) Cooperon ladders pick up an extra $\mp 2i\Delta$ as
described above; (c) the Hikami box in $\delta P_1$ picks up an extra
$\mp 2i\Delta$, whilst the box in $\delta P_3$ is unaffected. This
leads to the formulas
%%%%%%%%%%%%%%%%%%%%%%%%%%%%%%%%%%%%%%%%%
\begin{eqnarray}
{\delta P_1\over N(0)}&=&-{t\over 2}\Re e \sum_{m=1}^{M} \left\{
2\left[\psi\left({1\over 2}+m+\kappa\right)-
\psi\left({1\over 2}+\kappa\right)\right]I_2(m)+
\left[\psi'\left({1\over 2}+\kappa\right)-
\psi'\left({1\over 2}+m+\kappa\right)\right]I_1(m)\right\}\nonumber\\
{\delta P_2\over N(0)}&=&+{t\over 2}\Re e \sum_{m=1}^{M}
\left[\psi'\left({1\over 2}+\kappa\right)-
\psi'\left({1\over 2}+m+\kappa\right)\right]\overline{C}_1(m)\nonumber\\
{\delta P_3\over N(0)}&=&-t\Re e \sum_{m=1}^{M} {1\over m+2\kappa}
\left[\psi\left({1\over 2}+m+\kappa\right)-
\psi\left({1\over 2}+\kappa\right)\right](I_1(m)+2\alpha I_2(m))\nonumber\\
{\delta P_4\over N(0)}&=&-t\Re e \sum_{m=1}^{M} {1\over m+2\kappa}
\left[\psi\left({1\over 2}+m+\kappa\right)-
\psi\left({1\over 2}+\kappa\right)\right]\overline{C}_1(m)
\end{eqnarray}
%%%%%%%%%%%%%%%%%%%%%%%%%%%%%%%%%%%%%%%%%
where $I_1(m)$ and $I_2(m)$ are the same as before, whilst
%%%%%%%%%%%%%%%%%%%%%%%%%%%%%%%%%%%%%%%%%
\begin{equation}
\overline{C}_1(m)=
{4\pi D\over aL}\sum_{q} {1\over Dq^2+2\pi Tm-2i\Delta}
={L_T^2\over \pi aL}\sum_{n=-n_0}^{n_0}\sum_{k=0}^{k_0}
{1\over\left[m+2\kappa+
\left({kL_T\over 2a}\right)^2+\left({nL_T\over L}\right)^2
\right]}
\end{equation}
%%%%%%%%%%%%%%%%%%%%%%%%%%%%%%%%%%%%%%%%%
Calculating the $\delta P_{1-4}$ as above, or with $\beta$ set to
zero, we find essentially no difference over a wide range of magnetic
fields until we get to widths of order $10\AA$. It follows that we
only need consider the Zeeman splitting at the mean field level. Even
this is a small effect, only noticeable for wires narrower than about
$150\AA$.

Let us finally perform calculations of $T_c$ using the resummation
technique.  From the discussions above, we now know that to include
magnetic field effects we need only replace $Dq^2$ in Cooperons by
$2/\tau_H$, which must be calculated numerically, and include the
Zeeman splitting at the mean field level. The Cooperon denominator
$Dq^2+2|\epsilon_m|$ then becomes $2|\epsilon_m|+2/\tau_H-2i\Delta$,
and hence ${\hat\epsilon}_{nm}=(n+1/2+\kappa)\delta_{nm}$, which is
now complex. Because the two spin configurations which occur in an
s-wave pair propagator are complex conjugates of each other, we end up
taking the real part of the complex matrix equation.  The equation for
$\Lambda_{nm}$ becomes
%%%%%%%%%%%%%%%%%%%%%%%%%%%%%%%%%%%%%%%%
\begin{equation}
\Lambda_{nm}=
\cases{
I_1(n+m+1+2\alpha) &\quad\quad$\epsilon_n\epsilon_m>0$\cr
\displaystyle I_1(n+m+1+2\alpha)+2I_2(n+m+1+2\alpha)
&\quad\quad$\epsilon_n\epsilon_m<0$\cr}
\end{equation}
%%%%%%%%%%%%%%%%%%%%%%%%%%%%%%%%%%%%%%%%
where we have not included the Zeeman effect as we have shown this to be much 
smaller in perturbation theory.

We are now in a position to calculate the transition temperature,
$T_c$, as a function of resistance per square, $R_{\square}$, wire
width, $a$, and applied magnetic field, $H$. This obviously leads to a
wide choice of ways to present results. Experimentally one would
probably vary magnetic field on a sample of fixed width and thickness,
and measure $T_c(H)$. In Fig. 5 we therefore plot $T_c(H)$ for wires
of different resistances per square, $R_{\square}=0\Omega, 1000\Omega,
2000\Omega$, and widths $a=500\AA, 600\AA, 750\AA, 2D$ film. There is
far too much data on this plot, but some general trends are
apparent. Three physical processes are occuring here: increasing
$R_{\square}$ increases the effective Coulomb repulsion which strongly
suppresses $T_c$; decreasing $a$ means that a larger magnetic field is
required to fit a flux quantum inside the wire, and hence $H_{c2}$
increases at fixed $R_{\square}$; decreasing $a$ also causes a
crossover from two to one-dimensional behavior, and so increases
fluctuation effects, leading to a small reduction of $T_c$ at fixed
$R_{\square}$. The last two effects mean that the $T_c(H)$ curves for
wires with the same $R_{\square}$ and different $a$ cross each other
at non-zero field.

Another effect of the dimensional crossover is to change the shape of
the $T_c(H)$ curve.  This occurs because the magnetic pair-breaking
rate, $1/\tau_H$, is linear in magnetic field in the two-dimensional
limit, but quadratic in magnetic field in the one-dimensional
limit. This leads to two limiting forms for the dimensionless $T_c(H)$
curve,
%%%%%%%%%%%%%%%%%%%%%%%%%%%%%%%%%%%%%%%%%%%%
\begin{equation}
\label{tchlimits}
\ln{\left({T_c(H)\over T_c(0)}\right)}=\psi\left({1\over 2}\right)-
\psi\left({1\over 2}+e^{\psi(1/2)}\left[{H\over H_{c2}}\right]^p
\left[{T_{c}(0)\over T_c(H)}\right]\right),
\end{equation}
%%%%%%%%%%%%%%%%%%%%%%%%%%%%%%%%%%%%%%%%%%%%
where $p=1$ in the 2D limit and $p=2$ in the 1D limit. We can move
from the 2D limit to the 1D limit either by decreasing $a$ at fixed
$R_{\square}$, as shown in Fig. 6, or by increasing $R_{\square}$ at
fixed $a$, as shown in Fig. 7. Considering first the width dependence,
we see in Fig. 6 that the dimensional crossover occurs when $a\sim
L_T$. For the wider wires the $T_c(H)$ curve joins the two-dimensional
curve at the value of the magnetic field such that the magnetic
length, $L_H$, is of order the width, $a$; for narrower wires this
value of $H$ exceeds the two-dimensional value of $H_{c2}$, and the
curves never meet. For ultranarrow wires, $a<250\AA$, the Zeeman
effect becomes important. This causes the $T_c(H)$ curves to change
shape again, moving slowly back in the direction of the 2D limit.  Now
let us consider the resistance dependence of the $T_c(H)$ curve.  For
narrow wires, $a\approx 250\AA$, all $T_c(H)$ curves fall on the 1D
curve independent of $R_{\square}$. For wide wires, $a\approx
10000\AA$, all $T_c(H)$ curves fall on the 2D curve independent of
$R_{\square}$. For wires of intermediate width, the $T_c(H)$ curves
can be changed from the 1D curve to the 2D curve by increasing
$R_{\square}$, as shown in Fig. 7.  Finally for ultranarrow wires,
$a\approx 50\AA$, the $T_c(H)$ curves can again be tuned by varying
$R_{\square}$, in this case the tuning being between one-dimensional
orbit and Zeeman curves.

\section{Discussion and Conclusions}

In this paper we have considered the effect of localization and
interaction on transition temperature in ultranarrow superconducting
wires. Our main conclusion is that, taking into account the relatively
small amount of experimental data, theory and experiment are
consistent with each other.  The agreement is not perfect: the
experimental data seems to show a very sudden crossover from 2D to 1D
behavior when the wire width becomes of the order of the thermal
length, whilst theory predicts a smooth crossover.  Indeed, when
plotted as the additional relative suppression of transition
temperature due to finite width, the experimental data falls on two
curves -- one for the 2D film and $1000\AA$ wire; the other for the
$580\AA$, $350\AA$ and $250\AA$ wires. These conclusions are based on
a single experiment\cite{XHD97}, as this is the only experiment which
shows any consistent trend with wire width. More experimental data is
obviously needed to investigate this further. Given the difficulty in
fabricating these wires, we suggest that the magnetic field dependence
of transition temperature be measured in future experiments. We have
therefore made predictions for magnetic field dependence of transition
temperature in wires of varying width and resistance in the hope that
such measurements will be made in the future. Our main prediction is
of a change in the shape of the reduced $T_c(H)$ curves due to the
dimensional crossover. This may be tuned either by varying wire width
or resistance.

Our work is based on the resummation technique developed by Oreg and
Finkel'stein\cite{OF99}, which is a very powerful method of going
beyond perturbation theory, and hence giving believable results at
strong disorder. The suppression of the transition temperature is very
sensitive to boundary conditions, due to the different relative
weights of the zero momentum mode. We have used the physically correct
zero-gradient boundary conditions in our work. The unexpected strong
dependence of results on boundary conditions led us to worry about all
details of the calculation. We therefore also investigated the effects
of using a featureless Coulomb interaction instead of the exact one,
ignoring self-energy contributions, and solving the OF matrix equation
exactly instead of using an approximate method. We found that it is
legitimate to use a featureless Coulomb interaction and ignore
self-energy contributions, but that our matrix equations should be
solved by exact diagonalization (which in any case is no more
difficult).

In our calculations of the magnetic field dependence of transition
temperature, we find that the orbital effect is the most important,
and can be treated by replacing the $Dq^2$ terms in the momentum
dependent pair propagator, $L(q,0)$, by twice the magnetic dephasing
rate, $2/\tau_H$. The latter is the lowest eigenvalue of the minimally
coupled Cooperon operator, and must be found numerically. The Zeeman
splitting is found to have a non-negligible effect only at the mean
field level, and even this effect is small. The perturbative term only
becomes important in very narrow wires (of order $10\AA$), which
cannot currently be fabricated.

\bigskip
\centerline {\bf ACKNOWLEDGEMENTS}
\medskip

We thank I. Aleiner, A. Clerk, A.M. Finkel'stein, I.V. Lerner, Y. Oreg
and M. Steiner for helpful discussions, and F. Sharifi and P. Xiong
for making their data available to us.  RAS acknowledges support from
the Nuffield Foundation and the UK EPSRC under Grant No GR/M98975.
BSH acknowledges support from the UK EPSRC.  VA is supported by the US
National Science Foundation under Grant No DMR-9805613.

\medskip
\appendix
\section{Details of Some Momentum Sums}
\medskip

In this appendix we derive expressions for the quantitities $I_1(m)$
and $I_2(m)$ defined in Eqn. (\ref{Idef}) as
%%%%%%%%%%%%%%%%%%%%%%%%%%%%%%%%%%%%%%%%%%%%%%%%%%%%%%%%%%%%%%%%%%%%%%%%%%%%%%%%
\begin{eqnarray}
\label{AIdef}
I_1&\equiv& {4\pi D\over aL}\sum_{q^{\prime}} \frac{1}{Dq^{\prime 2}
+|\omega_m|}\nonumber\\ I_2&\equiv& {4\pi D\over aL}\sum_{q^{\prime}}
\frac{2\pi T}{(Dq^{\prime 2} +|\omega_m|)^2}
\end{eqnarray}
%%%%%%%%%%%%%%%%%%%%%%%%%%%%%%%%%%%%%%%%%%%%%%%%%%%%%%%%%%%%%%%%%%%%%%%%%%%%%%%%
In two dimensions the sum becomes an integral, and cutting off in a circularly
symmetric manner, 
%%%%%%%%%%%%%%%%%%%%%%%%%%%%%%%%%%%%%%%%%%%%%%%%%%%%%%%%%%%%%%%%%%%%%%%%%%%%%%%%
\begin{equation}
\label{2Dintdef}
\frac{1}{aL}\sum_{q^{\prime}}=\int_0^{Dq^{\prime 2}=1/\tau}\frac{2 \pi
q^{\prime} dq^{\prime}}{(2 \pi)^2} =\int_{0}^{1/\tau}
\frac{d(Dq^{\prime 2})}{4\pi D},
\end{equation}
%%%%%%%%%%%%%%%%%%%%%%%%%%%%%%%%%%%%%%%%%%%%%%%%%%%%%%%%%%%%%%%%%%%%%%%%%%%%%%%%%
we obtain the results
%%%%%%%%%%%%%%%%%%%%%%%%%%%%%%%%%%%%%%%%%%%%%%%%%%%%%%%%%%%%%%%%%%%%%%%%%%%%%%%%%
\begin{equation}
\label{2DIs}
I_1(m)=\ln{\left({M\over m}\right)},\quad
I_2(m)={1\over m}.
\end{equation}
%%%%%%%%%%%%%%%%%%%%%%%%%%%%%%%%%%%%%%%%%%%%%%%%%%%%%%%%%%%%%%%%%%%%%%%%%%%%%%%%% 
In quasi-1D wires we replace the integral in the transverse direction by a sum,
so that $I_1(m)$ becomes
%%%%%%%%%%%%%%%%%%%%%%%%%%%%%%%%%%%%%%%%%%%%%%%%%%%%%%%%%%%%%%%%%%%%%%%%%%%%%%%%%
\begin{eqnarray}
\label{1DI1}
I_1(m)&=&4\pi D\frac{1}{a} \sum_{q^{\prime}_a} \frac{1}{2 \pi}
\int_0^{Dq^{\prime 2}_L=1/\tau}
 \frac{ dq^{\prime}_L}{Dq^{\prime 2}_L + Dq^{\prime 2}_a +2 \pi T m}
 \nonumber \\ &=& \frac{4}{a} \sum_{q^{\prime}_a}
\frac{1}{\displaystyle\sqrt{q^{\prime 2}_a + \frac{2 \pi T m}{D}}}
\tan^{-1} \left[ \frac{1/\sqrt{D \tau}}
{\displaystyle\sqrt{q^{\prime 2}_a + \frac{2 \pi T m}{D}}} \right],
\end{eqnarray}
%%%%%%%%%%%%%%%%%%%%%%%%%%%%%%%%%%%%%%%%%%%%%%%%%%%%%%%%%%%%%%%%%%%%%%%%%%%%%%%%%%%%%%%%
and $I_2(m)$ can be found by differentiating the above with respect to $m$ and dividing
by $-2\pi T$. Substituting in the two types of boundary condition listed in Table 1
yields
%%%%%%%%%%%%%%%%%%%%%%%%%%%%%%%%%%%%%%%%%%%%%%%%%%%%%%%%%%%%%%%%%%%%%%%%%%%%%%%%%%%%%%
\begin{eqnarray}
I_1(m)=\cases{
\displaystyle  \frac{2}{\pi}\sum_{p=-p_0}^{p_0} (p^2 +mA^2)^{-\frac{1}{2}}
\tan^{-1}{\left[ \frac{p_0}{\sqrt{p^2 +mA^2}} \right]}&\quad\hbox{PBC}\cr
\displaystyle  \frac{4}{\pi}\sum_{p=0}^{2p_0} (p^2 +4mA^2)^{-\frac{1}{2}}
\tan^{-1}{\left[ \frac{2p_0}{\sqrt{p^2 +4mA^2}} \right]}&\quad\hbox{ZGBC}\cr}
\end{eqnarray}
%%%%%%%%%%%%%%%%%%%%%%%%%%%%%%%%%%%%%%%%%%%%%%%%%%%%%%%%%%%%%%%%%%%%%%%%%%%%%%%%%%%%%%
and
%%%%%%%%%%%%%%%%%%%%%%%%%%%%%%%%%%%%%%%%%%%%%%%%%%%%%%%%%%%%%%%%%%%%%%%%%%%%%%%%%%%%%%
\begin{eqnarray}
I_2(m)=\cases{
\displaystyle \frac{A^2}{\pi}\sum_{p=-p_0}^{p_0} (p^2 +mA^2)^{-\frac{3}{2}}
\left( \tan^{-1} \left[ \frac{p_0}{\sqrt{p^2 +mA^2}} \right] +
\frac{p_0 \sqrt{p^2 +mA^2}}{p^2+(m+M)A^2} \right) &\quad${\rm PBC}$\cr
\displaystyle \frac{8A^2}{\pi} \sum_{p=0}^{2p_0} (p^2 +4mA^2)^{-\frac{3}{2}}
\left( \tan^{-1} \left[ \frac{2p_0}{\sqrt{p^2 +4mA^2}} \right] +
\frac{2p_0 \sqrt{p^2 +4mA^2}}{p^2+4(m+M)A^2} \right) &\quad${\rm ZGBC}$\cr}
\end{eqnarray}
%%%%%%%%%%%%%%%%%%%%%%%%%%%%%%%%%%%%%%%%%%%%%%%%%%%%%%%%%%%%%%%%%%%%%%%%%%%%%%%%%%%%%%%
where $A=a/L_T$ is a dimensionless width, the ratio of the wire width
to the thermal length.

\newpage

%%%%%%%%%%%%%%%%%%%%%%%%%%%%%%%%%%%%%%%%%%%%%%%%%%%%%%%%%%%%%%%%%%%%%%%%%%%%%%%%%%%%%%%%%%
\begin{table}
\begin{center}
\begin{tabular}{|c|c|c|c|}
Type & Definition &\hspace{0.5cm} Momenta, \mbox{$q_a$}
\hspace{0.5cm}&\hspace{0.5cm}Details of sum over \mbox{$p$}\hspace{0.5cm} \\
\hline
&&&\\ Periodic & \mbox{$\psi(x+a)=\psi(x)$} & \mbox{$\displaystyle{2\pi p\over a}$}
& \mbox{$\displaystyle\sum_{p=-p_0}^{p_0}$}\\ &&&\\
\hline
&&&\\ Zero-Gradient &\hspace{0.5cm} \mbox{$\psi'(x=\pm a/2)=0$}\hspace{0.5cm}
& \mbox{$\displaystyle{\pi p\over a}$} & \mbox{$\displaystyle\sum_{p=0}^{2p_0}$}\\
&&& \\
%\hline
\end{tabular}
\end{center}
\end{table}
\null
\vskip -0.2truein
\noindent
{\ninerm
TABLE I. \ Details of the boundary conditions discussed in the text. The upper
cut-off in momentum is given by \mbox{$Dq_a^2=1/\tau$}, so that
\mbox{$p_0=a/2\pi\sqrt{D\tau}=\sqrt{M}(a/L_T)$} where \mbox{$L_T=\sqrt{2\pi D/T}$}
is the thermal length and \mbox{$M=1/2\pi T\tau$}. }
%%%%%%%%%%%%%%%%%%%%%%%%%%%%%%%%%%%%%%%%%%%%%%%%%%%%%%%%%%%%%%%%%%%%%%%%%%%%%%%%%%%%%%%%%%

\newpage

%%%%%%%%%%%%%%%%%%%%%%%%%%%%%%%%%%%%%%%%%%%
\begin{figure}[t]
\centerline{\psfig{figure=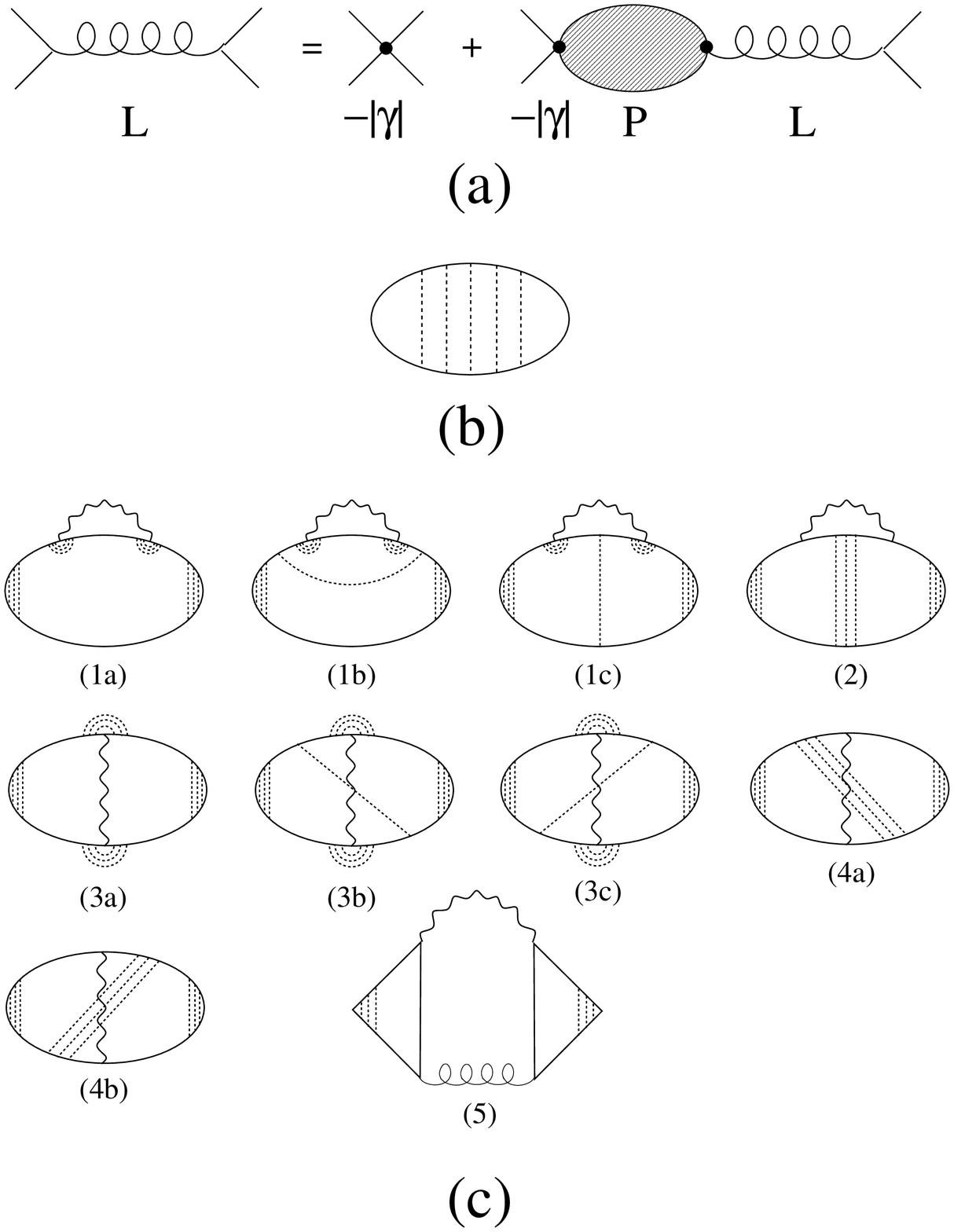,width=12cm}}
\end{figure}
\null
\vskip -0.2truein
\noindent
{\ninerm
FIG. 1. The diagrams needed to calculate the first-order perturbative
correction to the pair-propagator $L(q,0)$. (a) Definition of pair
propagator, $L(q,0)$, in terms of the pair polarization function,
$P(q,0)$, and the 4-point BCS interaction $-|\gamma|$. (b)
Zeroth-order (mean-field) polarization function, $P_0(q,0)$, in a
dirty superconductor. The solid lines are electron Green functions;
the dashed lines are impurity lines. (c) First-order perturbative
correction to pair polarization function, $\delta P(q,0)$. The wiggly
line is the screened Coulomb interaction.}
%%%%%%%%%%%%%%%%%%%%%%%%%%%%%%%%%%%%%%%%%%%%

\newpage

%%%%%%%%%%%%%%%%%%%%%%%%%%%%%%%%%%%%%%%%%%%%%%%%%%%%%%%%%%%%%%%%%%%%%%%%%%%%%%%%%%%%
\begin{figure}
\centerline{\psfig{figure=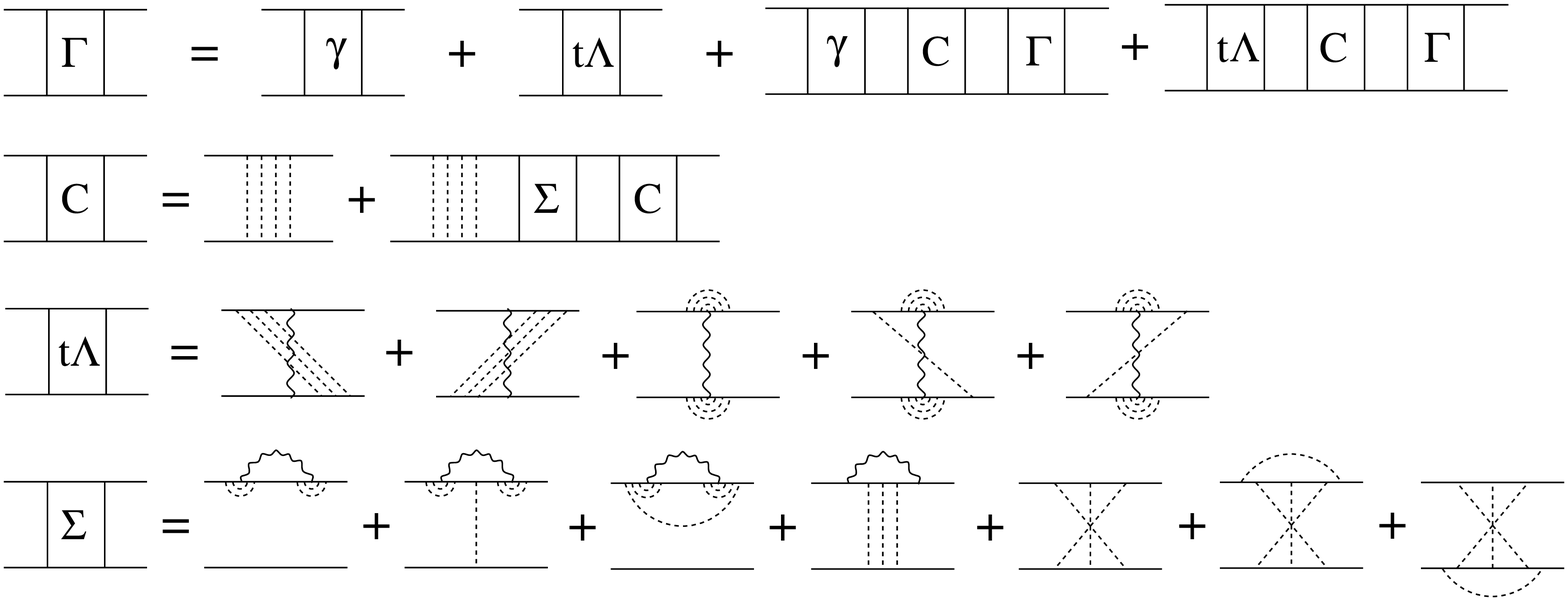,width=18cm}}
\end{figure}
\null
\vskip -0.2truein
\noindent
{\ninerm
FIG. 2. Diagrammatic equation for the pair amplitude matrix
$\Gamma(\epsilon_n,\epsilon_l)$. Block $\gamma$ is the BCS
interaction.  Block $t\Lambda$ is the correction to the effective
interaction caused by the interplay of Coulomb interaction and
disorder. Block $\Sigma$ is the correction to the bare Cooperon ladder
caused by Coulomb interaction and disorder, which yields a negligible
contribution and is thus ignored in our numerical calculations.}
%%%%%%%%%%%%%%%%%%%%%%%%%%%%%%%%%%%%%%%%%%%%%%%%%%%%%%%%%%%%%%%%%%%%%%%%%%%%%%%%%%%%

\newpage

%%%%%%%%%%%%%%%%%%%%%%%%%%%%%%%%%%%%%%
\begin{figure}[t]
\centerline{\psfig{figure=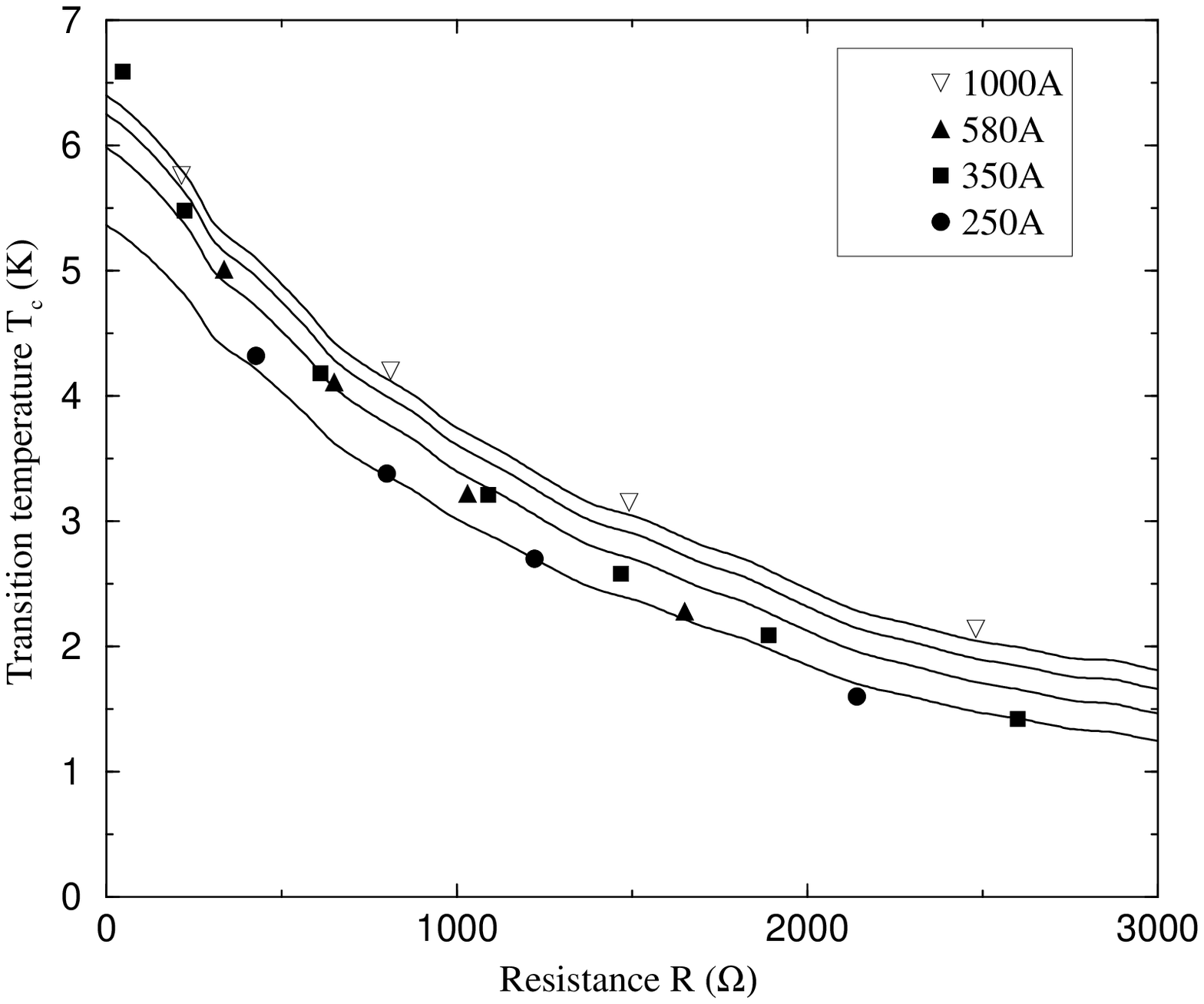,width=16cm}}
\end{figure}
\null
\vskip -0.2truein
\noindent
{\ninerm
FIG. 3. Comparison of theory (solid lines) and experiment (symbols) for
transition temperature, $T_c$, as a function of resistance per square,
$R_{\square}$, for various wire widths, $a=250\AA, 350\AA, 580\AA,
1000\AA$. The experimental data is that of Xiong et
al\cite{XHD97}. The theory used is the one we have argued to be
correct -- resummation technique solved by exact diagonalization using
zero-gradient boundary conditions. The theory plots are, from top to
bottom, for $1000\AA$, $580\AA$ $350\AA$ and $250\AA$ wires.
Rescaling has been applied to take into account the fact that theory
and experiment do not agree with each other in two dimensions, and
that the transition temperature at zero resistance appears to depend
upon width. The thermal length, $L_T$, has been taken to be $300\AA$
at $T_{c0}=6.9K$ throughout -- this is the value quoted by Sharifi et
al\cite{SHD93}. Plotting the data in this form focuses on the
suppression of transition temperature due to {\it disorder}, and we
see that the agreement between theory and experiment is good,
particularly for the $1000\AA$ and $250\AA$ wires.}
%%%%%%%%%%%%%%%%%%%%%%%%%%%%%%%%%%%%%%%

\newpage

%%%%%%%%%%%%%%%%%%%%%%%%%%%%%%%%%%%%%%
\begin{figure}[t]
\centerline{\psfig{figure=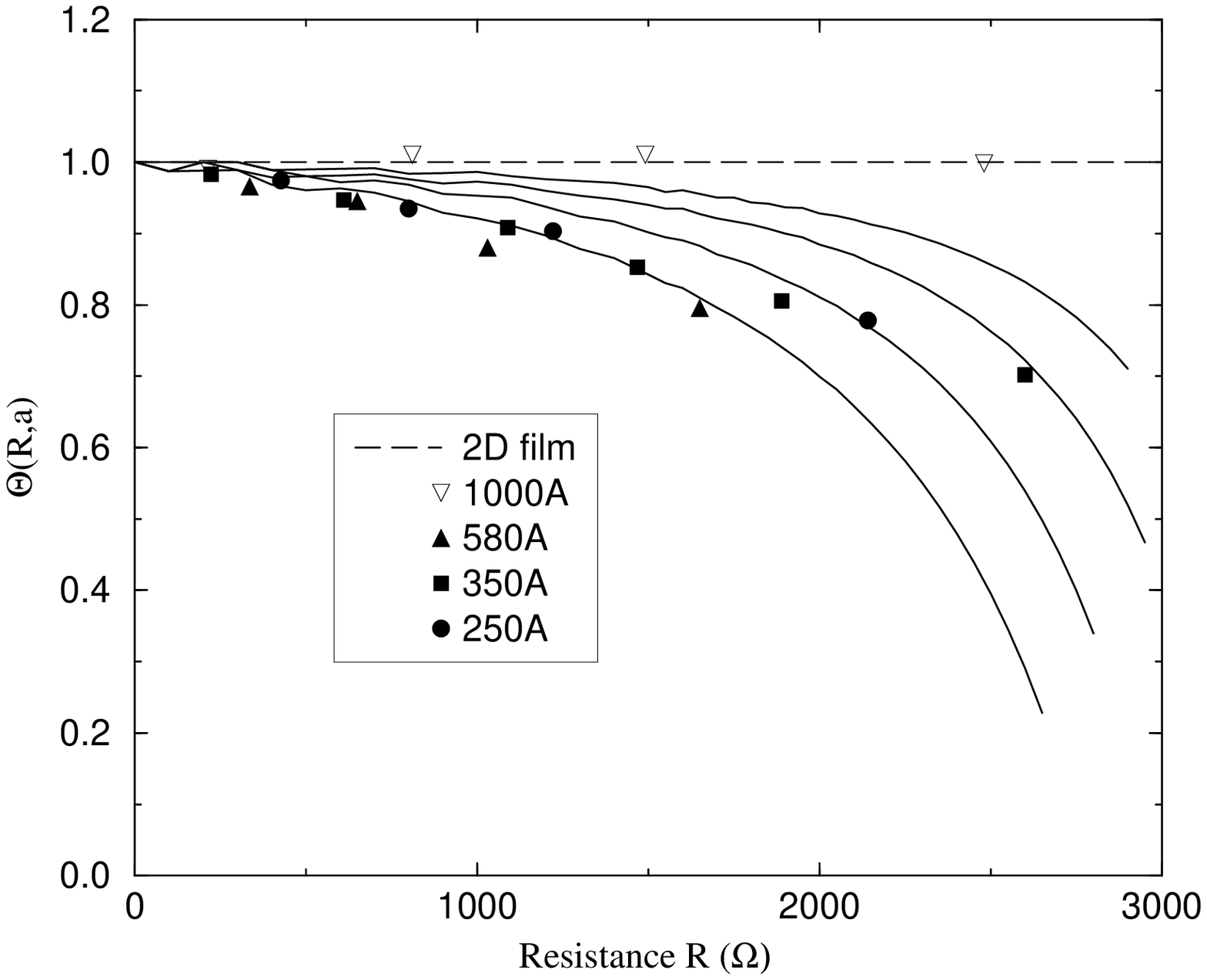,width=16cm}}
\end{figure}
\null
\vskip -0.2truein
\noindent
{\ninerm
FIG. 4. Comparison of theory (solid lines) and experiment (symbols) for the
additional relative suppression of transition temperature,
$\Theta(R_{\square},a)$, as a function of resistance per square,
$R_{\square}$, for various wire widths, $a=250\AA, 350\AA, 580\AA,
1000\AA$. Plotting the data in this form focuses on the suppression of
transition temperature due to {\it width}, and we see that the
agreement between theory and experiment is not particularly good. This
may have been anticipated from the experimental data which shows an
abrupt crossover between the 2D film and the $1000\AA$ wire, which lie
on one curve, and the $580\AA, 350\AA, 250\AA$ wires, which lie on a
different curve. Our theory, and indeed any theory, predicts a smooth
crossover.  }
%%%%%%%%%%%%%%%%%%%%%%%%%%%%%%%%%%%%%%

\newpage

%%%%%%%%%%%%%%%%%%%%%%%%%%%%%%%%%%%%%%
\begin{figure}[t]
\centerline{\psfig{figure=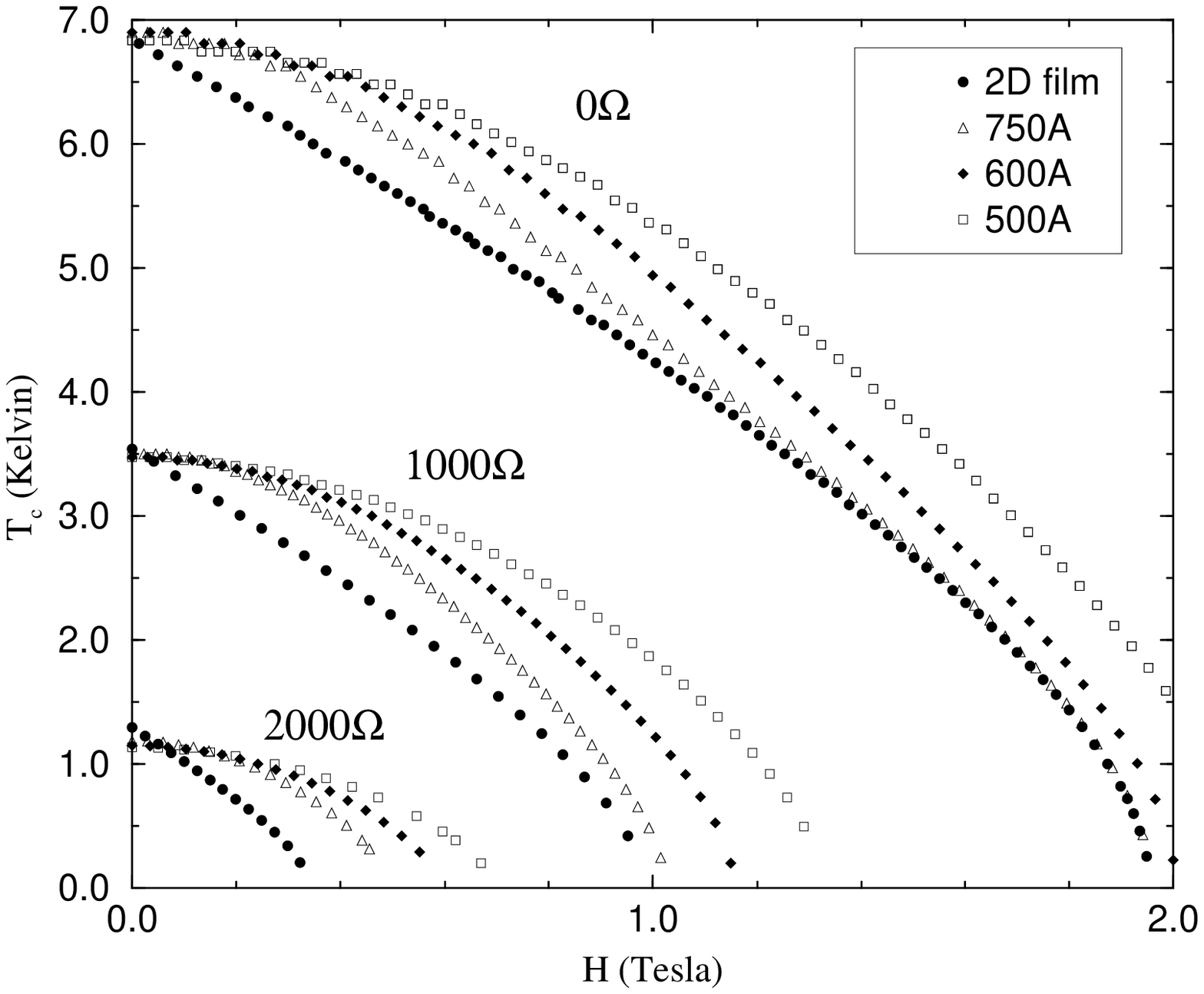,width=16cm}}
\end{figure}
\null
\vskip -0.2truein
\noindent
{\ninerm
FIG. 5. Predicted transition temperature, $T_c$, as a function of applied
perpendicular magnetic field, $H$, for wires of various widths,
$a=500\AA, 600\AA, 750\AA, 2D$ film, and resistances per square,
$R_{\square}=0\Omega, 1000\Omega, 2000\Omega$. Three physical effects
can be seen in this figure: increasing $R_{\square}$ reduces $T_c$
dramatically due to the increased Coulomb repulsion; decreasing $a$
increases the upper critical field, $H_{c2}$, as a larger field is
needed to fit a flux quantum into the wire; decreasing $a$ at fixed
$R_{\square}$ causes a small additional suppression of $T_c$ at zero
field due to increased fluctuation effects caused by the crossover
from two to one-dimensional behavior. The last two effects mean that
the $T_c(H)$ curves for wires with the same $R_{\square}$ and
different $a$ cross each other at non-zero field. Finally the shape of
the $T_c(H)$ curve changes as width $a$ is altered due to the
dimensional crossover. There is obviously a large amount of data in
this figure, and to analyse the individual trends we will need to
re-plot some of this data in a different form.  }
%%%%%%%%%%%%%%%%%%%%%%%%%%%%%%%%%%%%%%%%

\newpage

%%%%%%%%%%%%%%%%%%%%%%%%%%%%%%%%%%%%%%
\begin{figure}[t]
\centerline{\psfig{figure=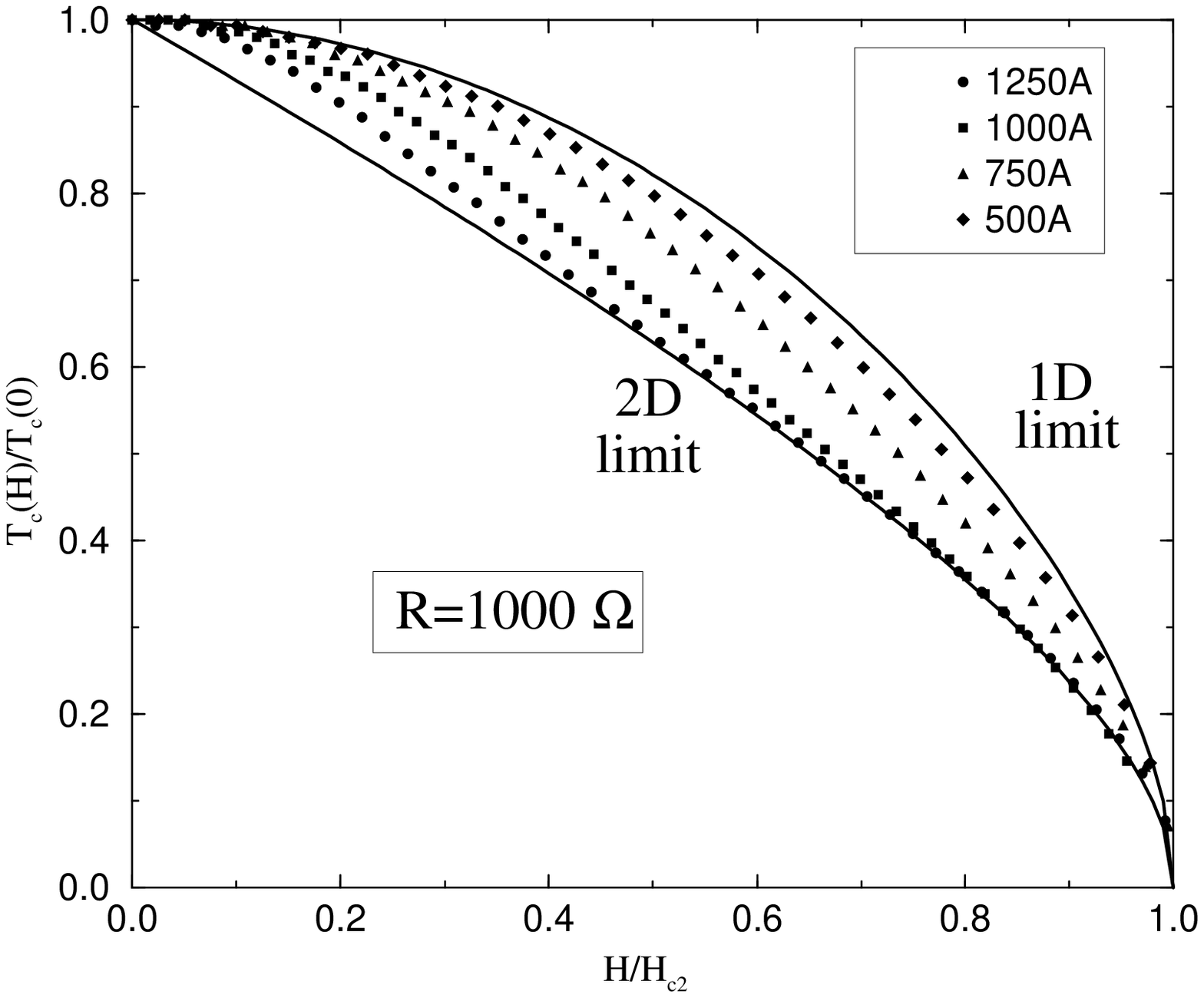,width=16cm}}
\end{figure}
\null
\vskip -0.2truein
\noindent
{\ninerm
FIG. 6. Predicted dependence of the shape of the $T_c(H)$ curve on wire width,
$a$, at a fixed resistance per square, $R_{\square}=1000\Omega$. We
plot using dimensionless variables $T_c(H)/T_c(H=0)$ vs $H/H_{c2}$ in
order to compare the curve shape. As width is reduced we see a
crossover from the two-dimensional limit, where the magnetic
pair-breaking rate $1/\tau_H\propto H$, to the one-dimensional limit
where $1/\tau_H\propto H^2$. The crossover occurs when the wire width
is of the order of the thermal length, $L_T$. For the wider wires the
$T_c(H)$ curve joins the two-dimensional curve at the value of the
magnetic field such that the magnetic length, $L_H$, is of order the
width, $a$; for narrower wires this value of $H$ exceeds the
two-dimensional value of $H_{c2}$, and the curves never meet. For
ultranarrow wires, $a<250\AA$, the Zeeman effect becomes
important. This causes the $T_c(H)$ curves to change shape again,
moving slowly back in the direction of the 2D limit.  }
%%%%%%%%%%%%%%%%%%%%%%%%%%%%%%%%%%%%%%%%

\newpage

%%%%%%%%%%%%%%%%%%%%%%%%%%%%%%%%%%%%%%
\begin{figure}[t]
\centerline{\psfig{figure=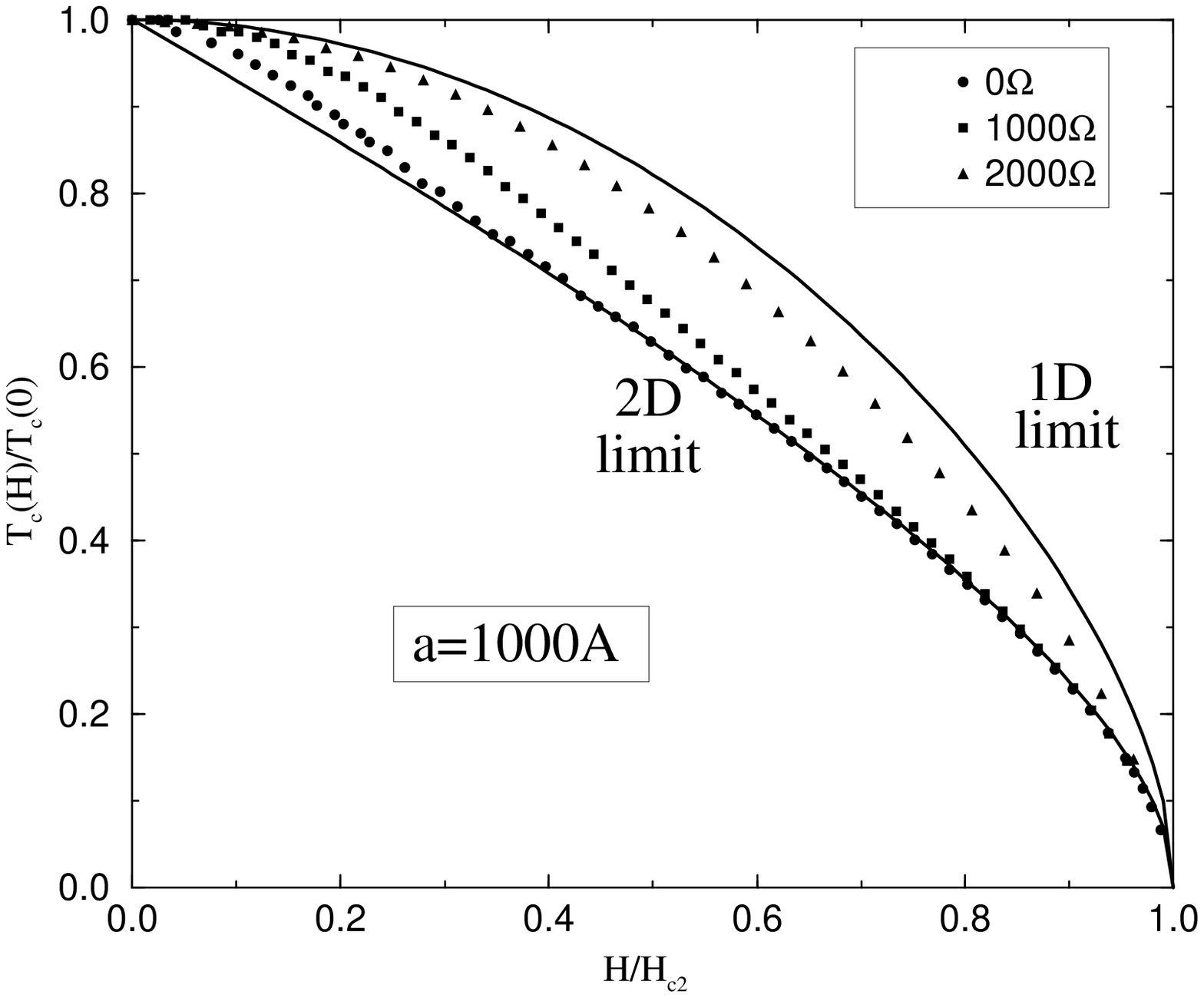,width=16cm}}
\end{figure}
\null
\vskip -0.2truein
\noindent
{\ninerm
FIG. 7. Predicted dependence of the shape of the $T_c(H)$ curve on resistance
per square, $R_{\square}$, at a fixed wire width, $a=1000\AA$. We see
a crossover from two-dimensional to one-dimensioanl behavior as
$R_{\square}$ is increased. This disorder-tuned crossover occurs only
for wires of intermediate width; a similar plot for $a=10000\AA$ shows
all data falling on the two-dimensional curve independent of
$R_{\square}$, whilst that for $a=250\AA$ shows all data falling on
the one-dimensional curve.  }
%%%%%%%%%%%%%%%%%%%%%%%%%%%%%%%%%%%%%%%% 

\begin{references}
\bibitem{OFBC} We note that Oreg and Finkel'stein\cite{OF99} also use
the physically correct zero-gradient boundary conditions, except that
their ratio $a/L_T$ is effectively half our value. This is unimportant
as the experimental value of $L_T$ is known only roughly
\cite{SHD93}. It follows that our predictions and theirs are in
agreement.
\end{references}
\end{document}